\documentclass[lettersize,journal]{IEEEtran}
\usepackage{amsmath,amsfonts,amssymb}
\usepackage{graphicx}
\usepackage[colorlinks]{hyperref}
\usepackage{changepage}
\usepackage{multicol,multirow,makecell}
\newcommand{\figref}[1]{Fig.~\ref{#1}}
\usepackage{mathtools}
\usepackage{subfigure}

\begin{document}
%\title{A Contemporary Survey on Fluid Antenna Systems:\\Fundamentals, Applications, and\\Networking Technologies}
\title{A Contemporary Survey on Fluid Antenna Systems:\\Fundamentals and Networking Perspectives}

\author{Hanjiang Hong,~\IEEEmembership{Member,~IEEE}, 
        Kai-Kit Wong,~\IEEEmembership{Fellow,~IEEE}, 
        Hao Xu,~\IEEEmembership{Senior Member,~IEEE},\\
        Xinghao Guo,
        Farshad Rostami Ghadi,~\IEEEmembership{Member,~IEEE},
        Yu Chen,~\IEEEmembership{Member,~IEEE},
        Yin Xu,~\IEEEmembership{Senior Member,~IEEE},\\
        Chan-Byoung Chae, \emph{Fellow, IEEE}, 
        Baiyang Liu,~\IEEEmembership{Senior Member,~IEEE},\\ 
        Kin-Fai Tong, \emph{Fellow, IEEE}, and 
        Yangyang Zhang

%        \vspace{-3mm}

\thanks{The work of K. K. Wong and F. Rostami Ghadi is supported by the Engineering and Physical Sciences Research Council (EPSRC) under Grant EP/W026813/1.}
\thanks{The work of H. Hong is supported by the Outstanding Doctoral Graduates Development Scholarship of Shanghai Jiao Tong University.}
\thanks{The work of H. Xu is supported by the Fundamental Research Funds for the Central Universities under grant 2242025R10001.}
\thanks{The work of Y. Xu and X. Guo is supported by National Natural Science Foundation of China Program (62371291,62422111).}
\thanks{The work of C.-B. Chae was in part supported by the Institute for Information and Communication Technology Planning and Evaluation (IITP)/NRF grant funded by the Ministry of Science and ICT (MSIT), South Korea, under Grant RS-2024-00428780 and 2022R1A5A1027646.}
\thanks{The work of K. F. Tong and B. Liu was funded by the Hong Kong Metropolitan University, Staff Research Startup Fund: FRSF/2024/03.}

\thanks{H. Hong, K. K. Wong, and F. Rostami Ghadi are with the Department of Electronic and Electrical Engineering, University College London, WC1E 7JE, London, United Kingdom. K. K. Wong is also affiliated with Yonsei Frontier Lab, Yonsei University, Seoul, South Korea (e-mail: $\rm\{hanjiang.hong,kai\text{-}kit.wong,f.rostamighadi\}@ucl.ac.uk$).}
\thanks{H. Xu is with the National Mobile Communications Research Laboratory, Southeast University, Nanjing 210096, China (e-mail: $\rm hao.xu@seu.ac.cn$).}
\thanks{X. Guo and Y. Xu are with the School of Information Science and Electronic Engineering, Shanghai Jiao Tong University, Shanghai 200240, China (e-mail: $\rm\{guoxinghao,xuyin\}@sjtu.edu.cn$).}
\thanks{Y. Chen is with the School of Information and Communication Engineering, Beijing University of Posts and Telecommunications, Beijing 100876, China (e-mail: $\rm yu.chen@bupt.edu.cn$).}
\thanks{C.-B. Chae is with the School of Integrated Technology, Yonsei University, Seoul, 03722 South Korea (e-mail: $\rm cbchae@yonsei.ac.kr$).}
\thanks{B. Liu and K. F. Tong are with the School of Science and Technology, Hong Kong Metropolitan University, Hong Kong SAR, China (e-mail: $\rm \{byliu, ktong\}@hkmu.edu.hk$).}
\thanks{Y. Zhang is with Kuang-Chi Science Limited, Hong Kong SAR, China (e-mail: $\rm yangyang.zhang@kuang\text{-}chi.org$).}

\thanks{Corresponding author: Kai-Kit Wong.}
}
\maketitle
\begin{abstract}
The explosive growth of teletraffic, fueled by the convergence of cyber-physical systems and data-intensive applications, such as the Internet of Things (IoT), autonomous systems, and immersive communications, demands a multidisciplinary suite of innovative solutions across the physical and network layers. {\color{blue}Fluid antenna systems (FAS) represent a transformative advancement in antenna design, offering enhanced spatial degrees of freedom through dynamic reconfigurability. By exploiting spatial flexibility, FAS can adapt to varying channel conditions, mitigate interference, and optimize wireless performance, making it a highly promising candidate for next-generation communication networks.} This paper provides a comprehensive survey of the state of the art in FAS research. We begin by examining key application scenarios in which FAS offers significant advantages. We then present the fundamental principles of FAS, covering channel measurement and modeling, single-user configurations, and the multi-user fluid antenna multiple access (FAMA) framework. Following this, we delve into key network-layer techniques such as quality-of-service (QoS) provisioning, power allocation, and content placement strategies. We conclude by identifying prevailing challenges and outlining future research directions to support the continued development of FAS in next-generation wireless networks.
\end{abstract}

\begin{IEEEkeywords}
Fluid antenna system (FAS), channel modeling, homogeneous network (HomoNet), networking, physical layer.
\end{IEEEkeywords}
%%%%%%%%%%%%%%%%%%%%%%%%%%%%%%%%%%%%%%%%%%%%%%%%%%%%%%%%%%%%%%%%%%%%%%%%%%%%%%%%%%%%%%%%%%%%%%%%%%%%%%%%%%%%%%%%%%%%%%%%%%%%%%%%%%%%%%%%%%%%%%%%%%%%%%%%%%%%%%%%

\section{Introduction}

\vspace{2mm}
\begin{quote}
\begin{center}
{\em ``Necessity is the mother of invention.''}
\end{center}
\end{quote}
\vspace{2mm}

\IEEEPARstart{F}{or wireless} communications, this truism is the ultimate {\em raison d'{\^e}tre}. From the inception of wireless telegraphy in the late 19th century to the contemporary era when the next generation networks serve as indispensable conduits between the digital and physical realms, the unyielding demand for faster, smarter, and more reliable communications has been a continual catalyst for innovation. 
While fifth-generation (5G) networks have achieved significant advancements in data transmission speed, latency, and connectivity, these enhancements are insufficient to accommodate the evolving demands of the cyber-physical world. The forthcoming sixth-generation (6G) network is expected to overcome the constraints of its predecessor, aiming to meet highly ambitious key performance indicators (KPIs). These benchmarks include a peak data rate of $1~{\rm Tbps}$, an end-to-end latency of $1~{\rm ms}$, and an extraordinary connection density of $10$ million devices per ${\rm km}^2$, etc~~\cite{I1tata20216Gwireless,I2_raja2020white,I3_saad2020avision,Tariq-2020}. Such objectives are not merely incremental upgrades. Rather, they are prerequisites for enabling the next generation of applications characterized by seamless interaction among humans, machines, and environments. 

To achieve these goals, researchers are investigating a multidisciplinary suite of innovative solutions. These include non-orthogonal multiple access (NOMA)~~\cite{Wang-2006,Saito-2013,Dai-2015}, rate-splitting multiple access (RSMA)~~\cite{Lin-2021,9831440}, extra-large multiple-output multiple-input (MIMO)~~\cite{I8_wang2024XLMIMO}, advanced coding and modulation schemes~~\cite{I37_rowshan2024channel,I36_hong2022enhanced,zhu2025transmissive,Li-2025}, and reconfigurable intelligent surfaces (RIS)~~\cite{Alexandropoulos-2021,wei2021channel,Basar-2024,xiao2024starris}. Researchers are increasingly focusing on the development of advanced networks driven by artificial intelligence (AI) and learning-based methods~~\cite{I6_letaief2019theroadmap,I5_wang2020artificial,I4_song2022networking}, Terahertz (THz) communications~~\cite{I10_aky2022THz,I11_chaccour2022THZ,I9_jiang2024THz}, non-terrestrial networks (NTNs)~~\cite{I13_araniti2022ntn,I12_mah2024NTN}, and integrated sensing and communication (ISAC)~~\cite{I17_liu2022isac,I18_liu2022isac,I16_lu2024isac}, etc. NOMA and RSMA utilizes advanced coding and decoding architectures to allow more spectrum sharing for greater capacity while RIS extends the principle of MIMO to be used at programmable metasurfaces in remote areas for passive but smart reflections. Nevertheless, scalability remains a critical limitation of these technologies. On the other hand, while AI is a powerful tool for optmization and resource allocation of wireless networks, AI itself is not a wireless communication technology. Shifting to the THz band is clearly a natural approach to access greater bandwidth, although the associated high cost poses a major challenge. The deployment of NTNs are application-specific. Evidently, ISAC makes communication more interesting but it also makes it a lot more difficult. In short, it remains essential to develop new technologies that introduce additional degree-of-freedom (DoF) at the physical layer, thereby expanding the achievable performance region, desperately needed for 6G.

To push the boundaries of physical-layer design, the fluid antenna system (FAS) stands out as a breakthrough technology, introducing dynamic reconfigurability in antenna position and structure, marking a leap in enhancing spatial DoF~~\cite{I19_wong2020fluid,I21_wong2022FAS}. Unlike conventional fixed-position antennas (FPA), a.k.a.~traditional antenna system (TAS), a FAS is a software-defined architecture that uses fluidic, conductive, or dielectric materials to facilitate real-time reconfiguration of antenna position and shape in a given spatial region. This innate adaptability constitutes an integral part of the FAS design, introducing additional DoFs that can be exploited to achieve significant performance enhancements. FAS was first conceptualized by Wong {\em et al.}~in 2020~~\cite{I19_wong2020fluid,I22_wong2020perflim,I20_wong2021FAS}, with its foundational idea partly inspired by Bruce Lee's philosophy of adaptability and flow:

\vspace{2mm}
\begin{quote}
\begin{center}
{\em ``Be like water making its way through cracks. Do not be assertive, but adjust to the object, and you shall find a way round or through it. If nothing within you stays rigid, outward things will disclose themselves. Empty your mind, be formless. Shapeless, like water. If you put water into a cup, it becomes the cup. You put water into a bottle and it becomes the bottle. You put it in a teapot it becomes the teapot. Now, water can flow or it can crash. Be water, my friend.''}
\end{center}
\end{quote}
\vspace{2mm}

{\color{blue}Applying this ``fluid'' philosophy to the physical layer of wireless communication leads to the conception of antennas that are inherently formless and highly adaptable, capable of dynamically reconfiguring themselves to cater to dynamic channel conditions and to exploit spatial DoF.} This idea is embodied in the FAS, which emphasizes extreme spatial reconfigurability and flexibility, marking a significant evolution in contemporary antenna design. The term `fluid' in FAS highlights the system's ability to rapidly adapt its configuration, without necessarily referring to liquid or gaseous states. Instead, it reflects the innovative, dynamic, and responsive nature of the architecture. FAS can be realized through various technological implementations, such as liquid-based antennas~\cite{I22_huang2021liquid, I23_paracha2019liquid, I24_shen2024design}, pixel-reconfigurable antennas~\cite{I25_hoang2021computational,I26_zhang2024pixel}, mechanically actuated antennas using stepper motors~\cite{I27_basbug2017design}, and flexible structures incorporating metamaterials~\cite{I28_johnson2015sidelobe,Deng-2023,Liu-2025arxiv}. In essence, FAS encompasses all forms of flexible and reconfigurable antennas that adhere to its underlying principle of spatial adaptability, which includes movable antennas~\cite{I30_zhu2024historical}, pinching antennas~\cite{Fukuda-NTT-22}, and the reconfigurable waveguide technology presented in~\cite{Wong-swc2021} when all the intelligent surfaces together combine to become an enormous FAS (E-FAS). Credit is also due to the foundational studies in~\cite{1367557,Fazel-2008,4200712}, which first explored reconfigurable antennas in the context of space-time coding and beamforming.

The motivations for this survey are multifaceted. In recent years, FASs have attracted growing attention due to significant theoretical progress and their potential for a wide range of applications. {\color{blue} As summarized in Table~\ref{tab-I-1}, }several review and tutorial articles have already explored various aspects of FAS. For example,~\cite{I21_wong2022FAS} identified six key research areas and discussed the promise of FAS under simplified channel models. In~\cite{I30_zhu2024historical}, the historical development of FAS and movable antennas was comparatively analyzed, while~\cite{I31_shoj2022MIMO} examined the interplay between FAS, MIMO, and RIS. Additional works, such as~\cite{I32_wong2023partI, I33_wong2023partII,I34_wong2023partIII}, have provided insights into the fundamental concepts of FAS, highlighted emerging research directions, and introduced a novel paradigm incorporating multiple RISs as distributed artificial scattering surfaces to facilitate massive connectivity. The study in~\cite{I36_shah2024asurvey} presented a comprehensive taxonomy of FAS applications in multiple access environments, termed fluid antenna multiple access (FAMA), and examined its potential integration with other cutting-edge technologies. More recently,~\cite{I35_New2024aTutorial} presented a comprehensive tutorial on the physical-layer foundations and enabling technologies of FAS. Lu {\em et al.}~in~\cite{Lu-2025} also offered an electromagnetic perspective to explain FAS. {\color{blue}~\cite{Shao20256DMA} presented a comprehensive taxonomy of a multi-dimensional reconfigurable FAS, namely six-dimensional movable antenna (6DMA).}
\begin{table*}
  \color{blue}
  \begin{center}
    \caption{Summary of Representative Survey Papers Related to FAS}\label{tab-I-1}
      \begin{tabular}{c|l}
        \hline
        \textbf{Reference} &  \textbf{Main Contributions} \\\hline
        \hline
        \cite{I21_wong2022FAS} & \makecell[l]{$\circ$ Identified six key research areas; \\ $\circ$ Discussed the promise of FAS under simplified channel models.}\\ \hline
        \cite{I30_zhu2024historical} &  $\circ$ Analyzed the historical development of FAS and movable antenna. \\ \hline
        \cite{I31_shoj2022MIMO} & $\circ$ Examined the interplay between FAS, MIMO, and RIS. \\ \hline
        \cite{I32_wong2023partI} & \makecell[l]{$\circ$ Introduced more recent channel models; \\ $\circ$ Provide theoretical performance of FAS and FAMA.} \\ \hline
        \cite{I33_wong2023partII} & $\circ$ Identified a few research opportunities in FAS. \\ \hline
        \cite{I34_wong2023partIII} & $\circ$ Investigated the use of RISs as distributed artificial scattering surfaces to facilitate massive connectivity.\\ \hline
        \cite{I36_shah2024asurvey} & \makecell[l]{$\circ$ Presented a comprehensive taxonomy of FAMA; \\ $\circ$ Examined the potential integration of FAMA with other technologies.}\\ \hline
        \cite{I35_New2024aTutorial} & \makecell[l]{$\circ$ Presented the fundamentals of FAS;\\ $\circ$ Introduced the strengths and weakness of different hardware designs; \\ $\circ$ Discussed the challenges and potential synergy of FAS.}\\ \hline
        \cite{Lu-2025} & $\circ$ Offered an electromagnetic explanation of FAS.\\ \hline
        \cite{Shao20256DMA} & $\circ$ Introduced a comprehensive taxonomy of multi-dimensional reconfigurable FAS, namely 6DMA. \\ \hline
        \hline
        This paper & \makecell[l]{$\circ$ Introduces the main scenarios and use applications of FAS in wireless communication networks;\\ $\circ$ Reviews the fundamentals of FAS, including the channel model, channel estimation methods, FAS, and FAMA;\\ $\circ$ Investigates the relevant network technologies of FAS for its integration into future 6G networks;\\ $\circ$ Identifies the new challenge and research directions of FAS.}\\ \hline
      \end{tabular}
  \end{center}
\end{table*}

Despite these valuable contributions, a significant gap remains in the literature concerning networking technologies specifically designed for FAS. This paper intends to address that gap by presenting a comprehensive survey of FAS, {\color{blue}with a particular focus on its integration into future 6G networks and the relevant networking techniques required to fully harness its potential. Specifically, we introduce its main scenarios and use applications in wireless communication networks, reviews its fundamentals, and focused on investigating the network technologies.} In doing so, this work aims to advance a holistic understanding of FAS and its transformative role in shaping next-generation wireless communication systems.

The reminder of this survey is organized as follows. Section~\ref{sec:app} introduces the main scenarios and use applications of FAS in wireless communication networks. Section~\ref{sec:fundamental} then discusses the fundamentals of FAS, including the measurement and modeling of FAS channels, as well as the physical technologies of single-user FAS and multi-user FAMA systems. In Section~\ref{sec:network}, we investigate relevant network techniques, such as quality of service (QoS) enhancement, power allocation, and content placement. Section~\ref{sec:challenges} identifies challenges and research directions, while Section~\ref{sec:conclusion} concludes the paper.  

%%%%%%%%%%%%%%%%%%%%%%%%%%%%%%%%%%%%%%%%%%%%%%%%%%%%%%%%%%%%%%%%%%%%%%%%%%%%%%%%%%%%%%%%%%%%%%%%%%%%%%%%%%%%%%%%%%%%%%%%%%%%%%%%%%%%%%%%%%%%%%%%%%%%%%%%%%%%%%%%
\section{Application Scenarios}\label{sec:app}
This section explores various application scenarios for FAS, highlighting the core operational characteristics that define their relevance. These application scenarios serve as a foundation for developing key enabling techniques tailored to FAS. Traditionally, most FAS deployments have been limited to homogeneous networks (HomoNets), characterized by macro-cell-only infrastructure. As shown in Table~\ref{TabIII_1_app}, these HomoNet scenarios can be broadly classified into two categories: Case~1, involving {\em single-user FAS networks}, and Case~2, involving {\em multi-user FAS networks}. More recently, FAS-assisted heterogeneous networks (HetNets) have gained attention, featuring integrated macro-cell and small-cell deployments {\color{blue} or cell-free network implementations}. The following sections delve deeper into each of these scenarios.

\begin{table*}
  \begin{center}
    \caption{Typical Application Scenarios}\label{TabIII_1_app}
    \begin{tabular}{c|c|l|l}
      \hline
      \multicolumn{2}{c|}{\textbf{Type}}  & \textbf{Description}  & \textbf{Features}\\ \hline\hline
      \multirow{5}{*}{\makecell[c]{Single-user\\HomoNet}}  & Case 1A & Basic FAS  & \makecell[l]{FAS on the transmitter and/or the receiver sides with one/multiple activated port(s);\\ $\circ$~Enhanced performance and increasing capacity.} \\\cline{2-4}
      % & Case 1B & MIMO-FAS  & \makecell[l]{FAS on the transmitter and the receiver sides with multiple activated ports;\\ Enhanced performance and increasing capacity.} \\\cline{2-4}
      & Case 1B & RIS-aided FAS & \makecell[l]{FAS on the receiver side; \\$\circ$~Effectively mitigating the double-path loss of RIS.} \\\cline{2-4}
      & Case 1C & FAS-aided IM  & \makecell[l]{Encoded FAS on the transmitter side;\\$\circ$~Improving spectral efficiency.}\\\hline
      \multirow{6}{*}{\makecell[c]{Multi-user\\HomoNet}}   & Case 2A & Multi-user MIMO-FAS & \makecell[l]{FAS on the transmitter and/or the receiver sides with one activated port;\\ $\circ$~Improving robustness and capacity.}\\\cline{2-4}
      & Case 2B & FAMA  & \makecell[l]{FAS on the receiver side;\\ $\circ$~Massive connectivity.}\\\cline{2-4}
      & Case 2C & FAS-aided ISAC & \makecell[l]{FAS on the transmitter side;\\$\circ$~Adaptive performance tuning;\\$\circ$~Improving spectral efficiency.} \\ \hline
      \multirow{5}{*}{HetNet} &  Case {\color{blue}3A} &  Content-centric FAS HetNets &  \makecell[l]{FAS on the receiver side;\\$\circ$~Enabling adaptive link establishment; \\$\circ$~Improving spectral efficiency;\\ $\circ$~Enhancing reliability and robustness.}\\ \cline{2-4}
      & \color{blue} Case 3B & \color{blue}  FAS-assisted Cell-free Networks & \color{blue} \makecell[l]{FAS on the access points or the user equipment;\\$\circ$~Improving spectral efficiency;\\$\circ$~Mitigating the interference; \\$\circ$~Alleviate the CSI overhead burden.}\\
      \hline
    \end{tabular}\vspace{-5mm}
  \end{center}
\end{table*}
%%%%%%%%%%%%%%%%%%%%%%%%%%%%%%%%%%%%%%%%%%%%%%%%%%%%%%%%%%%%%%%%%%%%%%%%%%%%%%%%

% \begin{figure}
%   \centering
%   \subfigure[]{\includegraphics[width =\linewidth]{Figure/SecII/FigII-A-1_BasicFAS.eps}\label{fig-basicFAS}}\\%\vspace{-3mm}
%   \subfigure[]{\includegraphics[width =\linewidth]{Figure/SecII/FigII-A-3_RIS-FAS.eps}\label{fig-RIS-FAS}}\\%\vspace{-3mm}
%   \subfigure[]{\includegraphics[width=\linewidth]{Figure/SecII/FigII-A-4_FA-IM.eps}\label{fig-FA-IM}}%\vspace{-2mm}
%   \caption{Illustration of (a) Case 1A: Basic FAS, (b) Case 1B: RIS-aided FAS, and (c) Case 1C: FAS-aided IM.} \vspace{-2mm}
% \end{figure}
\begin{figure*}
  \centering
  \includegraphics[width = 0.8\linewidth]{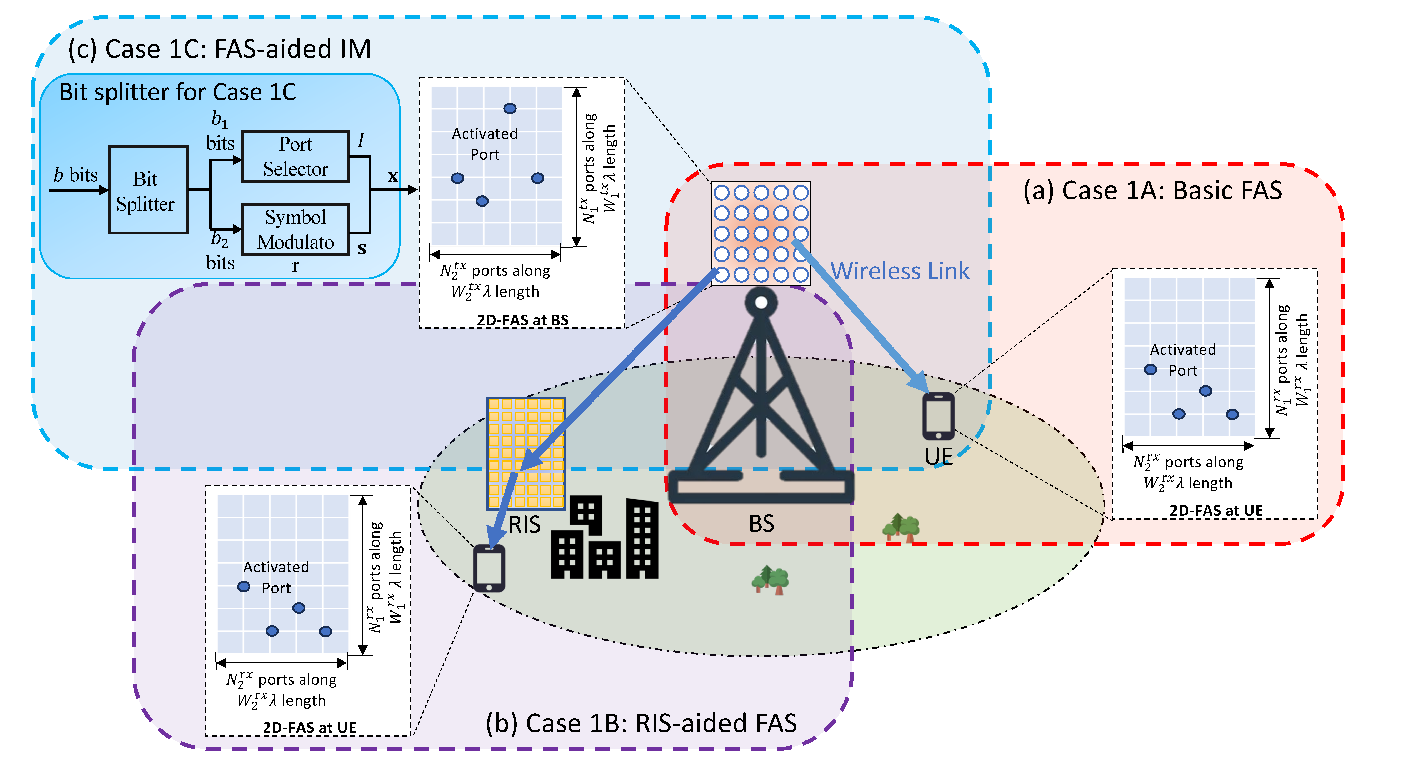}
  \caption{Illustration of (a) Case 1A: Basic FAS, (b) Case 1B: RIS-aided FAS, and (c) Case 1C: FAS-aided IM.} \label{fig-su-homo}\vspace{-2mm}
\end{figure*}

\subsection{Case 1: Single-user HomoNet FAS Scenarios} % Xinghao
\subsubsection{Case 1A: Basic FAS} %{fig-basicFAS}
In a basic FAS network, as shown in \figref{fig-su-homo}(a), FAS can be utilized at the transmitter and/or receiver. The basic FAS can be further categorized into two types based on the number of activated ports: single-input single-output FAS (SISO-FAS), which employs a single activated port, and MIMO-FAS, which utilizes multiple activated ports. For clarity, the subscript/superscript ${\rm s}$ is used to denote the parameters at the transmitter or receiver as ${\rm tx}$ or ${\rm rx}$, respectively, i.e., $ {\rm s} \in\{{\rm tx}, {\rm rx}\}$. In this case, \(N_i^{\rm s}\) ports are uniformly distributed along a linear segment of length \(W_i^\text{s} \lambda\), in which \(i\in \{1,2\}\), \(N_\text{s} = N_1^\text{s} \times N_2^\text{s}\) and \(W_\text{s} = W_1^\text{s} \lambda \times W_2^\text{s} \lambda\). If $N_2^\text{s} = 1$, the FAS configuration can be regarded as a one-dimensional linear FAS (1D-FAS); otherwise, if $N_i^\text{s} > 1,~\forall i \in \{1, 2\}$, the configuration is considered as a two-dimensional planar FAS (2D-FAS). FAS is switched to the port(s) with the strongest signal to enhance the receiving performance. 

Based on a simplified channel model, an early study in~\cite{I21_wong2022FAS} shows that the outage probability of SISO-FAS decreases with the number of ports, and FAS could outperform maximal-ratio combining (MRC) when the number of ports is sufficiently large. The effort in~\cite{I22_wong2020perflim} further derived the ergodic capacity and its lower bound, as well as the level crossing rate (LCR) and average fade duration (AFD). In~\cite{G3_chai2022SISO-FAS-PS}, an efficient port selection method for the SISO-FAS network that combines machine learning methods with analytical approximations, was proposed using only a few port observations. In~\cite{G5_new2023SISO-FAS}, the outage probability and diversity gain of SISO-FAS were derived using a simple yet accurate channel model that closely follows the spatial correlation of Jake's model, and an algorithm was proposed to approximate the value of the number of ports, $N^*$, at which the diversity gain saturates. Additionally, the performance of backscatter communication in SISO-FAS was analyzed in~\cite{G6_farshard2024SISO-FAS-BC}. Moreover, secure communication for a SISO-FAS network was investigated in~\cite{G7_farshard2024SISO-FAS-Secure}, where the key secrecy metrics, including secrecy outage probability, average secrecy capacity, and secrecy energy efficiency (EE) were analyzed.
% In~\cite{G4_zhu2024SISO-MA}, the deployment of the antenna position is investigated, considering a field-response channel model while omitting spatial correlation effects.

\begin{figure*}
  \centering
  \includegraphics[width = 0.75\linewidth]{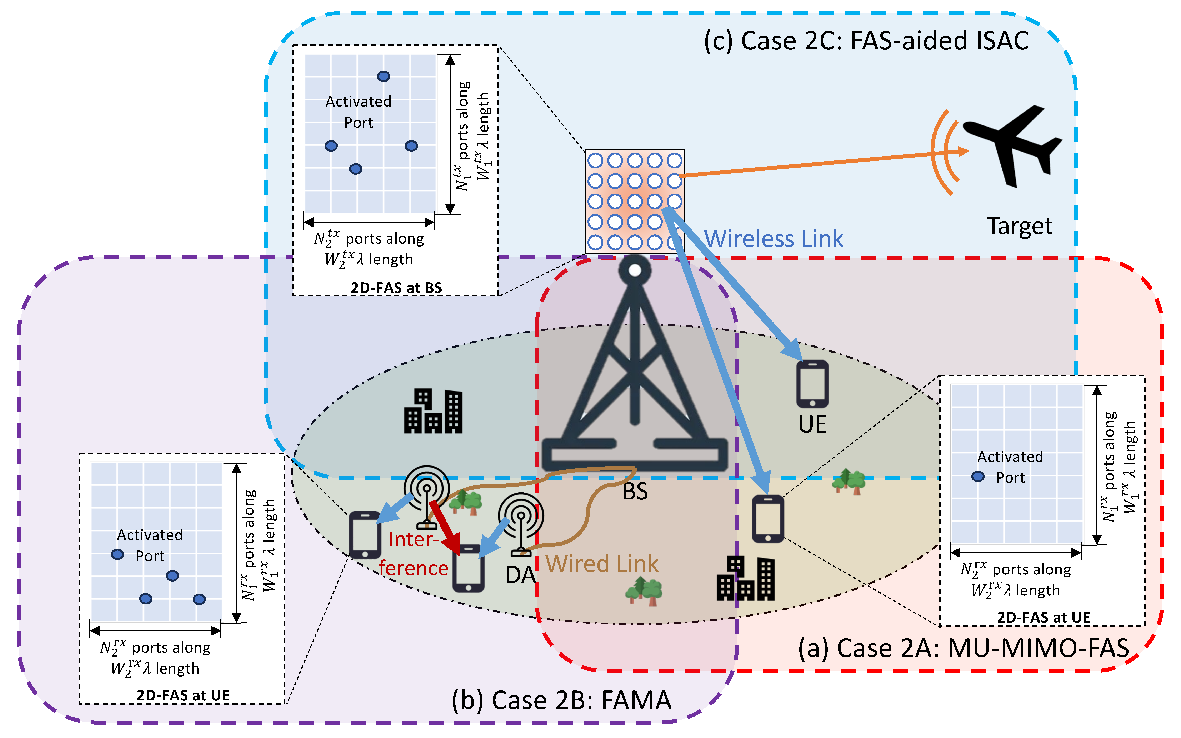}
  \caption{Illustration of (a) Case 2A: MU-MIMO-FAS, (b) Case 2B: FAMA, and (c) Case 2C: FAS-aided ISAC.}\vspace{-2mm}\label{fig_mu_homo}
\end{figure*}

Utilizing multiple fluid antennas at both ends of a point-to-point communication channel, referred to as MIMO-FAS, can provide enhanced performance compared with SISO-FAS and has recently been studied in~\cite{G10_new2024MIMO-FAS}. Specifically, the diversity-and-multiplexing trade-off (DMT) for the MIMO-FAS channel was characterized at high signal-to-noise ratio (SNR). Besides,~\cite{G11_ye2024MIMO-FAS} addressed the MIMO-FAS channel exploiting only statistical channel state information (CSI). Based on the rate maximization criterion, an algorithmic framework was proposed for transmit precoding and transmit/receive FAS position designs. Later in~\cite{G12_Krikidis2024MIMO-FAS}, the problem of antenna configuration selection for MIMO-FAS was studied using physics-inspired heuristics to maximize two fundamental performance metrics: the SNR at the receiver side and the end-to-end Shannon capacity. In~\cite{G13_Efrem2024MIMO-FAS}, the port selection problem on both the transmitter and receiver in MIMO-FAS was tackled to maximize the capacity. A new joint convex relaxation (JCR) problem is formulated, and two optimization algorithms are developed, each offering a different trade-off between performance and complexity.
% The study in~\cite{G9_ma2024MIMO-MA} indicates that the capacity of MIMO-FAS network can be improved by up to 30.3\% compared to traditional MIMO. 

\subsubsection{Case 1B: RIS-aided FAS}{fig-RIS-FAS}
RISs have emerged as a cost-effective technology for improving base station (BS) coverage by intelligently manipulating the propagation environment. By adjusting the phase shifts of their reflecting elements, RIS can steer radio signals toward intended receivers. When combined with FAS, RIS can produce synergistic performance gains. As shown in \figref{fig-su-homo}(b), the RIS-assisted FAS network comprises a BS equipped with a single FPA, a RIS with $\color{blue}M_{\rm RIS}$ reflecting elements, and a mobile user equipped with a FAS featuring $N_\text{rx}$ ports. In this scenario, the direct link between the BS and the user is obstructed by physical obstacles. As a result, the BS transmits the RF signal, which is redirected by the RIS to reach the mobile receiver. The FAS-assisted mobile receiver demonstrates superb performance, effectively mitigating the double-path loss in cascaded RIS systems~\cite{I34_wong2023partIII,G15_Ghadi2024FAS-RIS}. By applying the central limit theorem (CLT) and the block-correlation channel model,~\cite{G16_Lai2024FAS-RIS} simplified the expressions for outage probability, reduced computational complexity, and analyzed the impact of the number of FAS ports on system performance. In~\cite{G17_Yao2025FAS-RIS}, the authors further derived the upper bound, lower bound, and asymptotic approximation of the outage probability. Additionally, the analytical outage probability expression was used to design the passive beamforming of the RIS. A  framework for the RIS-aided FAS design was given in~\cite{G18_Yao2024FAS-RIS}, proposing two approaches: a CSI-based scheme and a CSI-free scheme. Then leveraging the derived outage probabilities,~\cite{G18_Yao2024FAS-RIS} optimized the throughput of the RIS-aided FAS. It is also possible to come up with a low-complexity beamforming design for RIS-aided systems if a fluid antenna array exploiting only statistical CSI is used at the BS~\cite{G19_Chen2024FAS-RIS}. Subsequently,~\cite{G20_farshard2024FAS-RIS-Secure} studied the impact of RIS-FAS for secure communication, in which an analytical expression for the secrecy outage probability was obtained. In~\cite{ghadi2025phase}, a simultaneously transmitting and reflecting (STAR)-RIS system with FAS under phase errors, was investigated and the benefits of position reconfigurability was highlighted. {\color{blue} More recently, the ``fluid'' concept of FAS was applied to the RIS elements and resulted in fluid reconfigurable intelligent surface (FRIS)~\cite{ye2025fris,salem2025fris,xiao2025fris}.}

\subsubsection{Case 1C: FAS-aided IM}{fig-FA-IM}
Index modulation (IM) leverages the indices of entities to encode and transmit information. FAS can be utilized as the encoded entities in IM to enhance the overall performance. As illustrated in \figref{fig-su-homo}(c), the FAS-aided IM network consists of the transmitter equipped with FAS, and the receiver equipped with $N_\text{rx}$ FPAs. The configuration of the FAS equipped at the transmitter is consistent with those in the aforementioned scenarios. The IM mechanism is applied to the fluid antenna ports, selecting $n_\text{tx}$ ports out of $N_\text{tx}$ ports for transmitting modulated symbols~\cite{G22_Faddoul2023IM-FAS, G25_Yang2024FAS-PIM, G27_Zhu2024FA-IM, G28_Guo2024FAG-IM}. Specifically, during each transmission interval, the input $b$-bit sequence is divided into two parts. The first part is mapped into $n_\text{tx}$ $\color{blue}2^{Q_m}$-ary constellation symbols, represented as $\boldsymbol{s} = [s_1, \dots, s_k, \dots, s_{n_\text{tx}}]^T$, while the second part is mapped to an index set of $n_\text{tx}$ ports, denoted as $\mathcal{I} = \{I_1, \dots, I_k, \dots, I_{n_\text{tx}}\}$, where $I_k \in \{ 1,2, \dots, N_\text{tx}\}$. Therefore, the length of the first part is calculated as $b_1 = n_\text{tx} {\color{blue}Q_m}$, {\color{blue}indicating that $n_\text{tx}$ parts of $Q_m$-bit sequence are mapping to $n_\text{tx}$ symbols,} while that of the second part is calculated as \( b_2 = \left \lfloor \log_2 \binom{N_\text{tx}}{n_\text{tx}} \right \rfloor \). The spectral efficiency of the FAS-aided IM network is therefore enhanced by the adequate possibility provided by the FAS port selection. In the initial studies, only one activated FAS port is considered for IM~\cite{G22_Faddoul2023IM-FAS}. As the study progresses, {\color{blue}multiple FAS port selection is discussed in~\cite{G25_Yang2024FAS-PIM} to implement IM}. To enhance spectral efficiency,~\cite{G27_Zhu2024FA-IM} applies the FAS-aided IM to MIMO systems, and~\cite{G28_Guo2024FAG-IM} further enhances the resilience of the systems in the spatial correlated channels. Also,~\cite{G23_Chen2024S-FAIM} proposes a secure FAS-aided IM framework that safeguards both index and data symbol transmissions against eavesdropping by employing a combined configuration of non-orthogonal spectrally efficient frequency division multiplexing waveforms and channel coding. Then~\cite{G24_Chen2024FAIM} combines FAS-aided IM with orthogonal frequency division multiplexing (OFDM) transmission systems and implements a wavelet scattering neural network in the proposed framework, achieving fast classification of index patterns using limited training data. A joint transmit and receive FAS-aided IM system enabled by RIS is also proposed in~\cite{G26_Zhu2024FAIM-RIS}. 
%%%%%%%%%%%%%%%%%%%%%%%%%%%%%%%%%%%%%%%%%%%%%%%%%%%%%%%%%%%%%%%%%%%%%%%%%%%%%%%%

%\vspace{-3mm}

% \begin{figure}
%   \centering
%   \subfigure[]{\includegraphics[width = \linewidth]{Figure/SecII/FigII-B-1_MUMIMOFAS.eps}\label{Fig_MIMO_FAS}}\\%\vspace{-2mm}
%   \subfigure[]{\includegraphics[width = \linewidth]{Figure/SecII/FigII-B-2_FAMA.eps}\label{Fig_FAMA}}\\%\vspace{-2mm}
%   \subfigure[]{\includegraphics[width = \linewidth]{Figure/SecII/FigII-B-3_FASaidedISAC.eps}\label{Fig_FAS_ISAC}}%\vspace{-2mm}
%   \caption{Illustration of (a) Case 2A: MU-MIMO-FAS, (b) Case 2B: FAMA, and (c) Case 2C: FAS-Aided ISAC.}\vspace{-2mm}
% \end{figure}

\subsection{Case 2: Multi-user HomoNet FAS Scenarios} % Hanjiang
\subsubsection{Case 2A: Multi-user MIMO-FAS}
Multi-user MIMO-FAS (MU-MIMO-FAS) extends from the concept of MIMO-FAS and enhances the traditional MU-MIMO by leveraging FAS's spatial reconfigurability. As illustrated in Fig.~\ref{fig_mu_homo}(a), the BS serves $U$ user equipments (UEs). Three scenarios can be considered: the first is to equip with a FAS with multiple RF chains at the BS and use a single traditional antenna with one RF chain on the UE side; the second is to use a multi-antenna uniform linear array (ULA) at the BS while each UE has a FAS; and the third is to adopt FAS on both sides. For the FAS at BS, $N_{\text{tx}} = N_1^{\text{tx}} \times N_2^{\text{tx}}$ antenna ports are uniformly distributed on a physical size of $W_\text{tx} = W_1^\text{tx}\lambda \times W_2^\text{tx}\lambda$. On the UE side, the distribution of the 2D-FAS can be denoted as $N_{\text{rx}} = N_1^{\text{rx}} \times N_2^{\text{rx}}$ ports over a size of $W_\text{rx} = W_1^\text{rx}\lambda \times W_2^\text{rx}\lambda$.
FAS equipped on both sides can reduce the impact of hardware imperfections through dynamic antenna repositioning, thereby improving the robustness in MU-MIMO systems~\cite{H1_yao2024rethinking}. Moreover, FAS on the UE side enables sparse channel recovery via compressed sensing, achieving lower estimation errors compared to traditional fixed antennas~\cite{xu2024channel}, and it can greatly improve the capacity of multiple access channel in the uplink system~\cite{H3_xu2024capacity}.

\subsubsection{Case 2B: FAMA}
FAMA leverages the unique feature of FAS to capitalize on spatial opportunities where interference is weak. As illustrated in Fig.~\ref{fig_mu_homo}(b), the BS serves $U$ UEs, each equipped with an $N_\text{rx} = N^\text{rx}_1\times N^\text{rx}_2$ ports FAS with a size of $W_\text{rx} = W^\text{rx}_1\lambda \times W^\text{rx}_2 \lambda$. All $U$ UEs reuse the same frequency radio resources. Thus, at each UE, interference channels with $(U-1)$ UEs are considered. The UE uses FAS to identify  and receive at the port with the weakest interference. Depending on how fast the UE updates the FAS port, FAMA can be roughly classified into \emph{fast} FAMA (\emph{f}-FAMA)~\cite{H4_wong2022FAMA,H5_wong2023fast} or \emph{slow} FAMA (\emph{s}-FAMA)~\cite{H6_wong2023sFAMA}. The \emph{f}-FAMA scheme switches the antenna port on a per-symbol basis, where the data-dependent sum interference plus noise signal cancels. This idea could achieve ultra-reliable massive access, supporting hundreds of users~\cite{H5_wong2023fast}. However, \emph{f}-FAMA could be impractical because of the complexity of instantaneously observing a large number of received signals. On the contrary, \emph{s}-FAMA is a more practical scheme because it only requires the FAS to switch the position once during each channel coherence time~\cite{H7_Espinosa2024Anew,H8_Xu2024revisiting}. Tens of users can be supported simultaneously by \emph{s}-FAMA. As an evolution of \emph{s}-FAMA, compact ultra massive antenna array (CUMA)~\cite{H12_Wong2024cuma} activates multiple antenna ports on the UE side to ensure that the received signals are combined constructively. Hence, the connectivity capability upgrades to hundreds of users again by CUMA~\cite{H13_wong2024transmitter}. Reinforcement learning is another promising solution to deal with the high-dynamic port selection~\cite{H9_Waqar2024Opportunistic}. Moreover, FAMA could be facilitated with coding and modulation to enhance its reliability~\cite{H10_hong2024coded,H11_hong2025Downlink,Ghadi2024fluid}. 

\subsubsection{Case 2C: FAS-aided ISAC}
FAS enhance ISAC by dynamically balancing sensing and communication trade-offs. As illustrated in Fig.~\ref{fig_mu_homo}(c), the BS is equipped with a FAS and accommodates $U$ users. The BS transmits common signals to the users and utilizes the echo signals to perform potential target sensing. For the FAS at BS, $N_{\text{tx}} = N_1^{\text{tx}} \times N_2^{\text{tx}}$ antenna ports are uniformly distributed on a physical size of $W_\text{tx} = W_1^\text{tx}\lambda \times W_2^\text{tx}\lambda$. FAS equipped at the BS can shift the ISAC Pareto frontier, enabling adaptive performance tuning~\cite{H16_Zou2024shifting}. In~\cite{H17_wang2024fluid}, reinforcement learning is used to optimize the antenna port selection and achieve $30\%$ gains on spectral efficiency. FAS can also be equipped on the user side to improve backscatter device detection and spectral efficiency~~\cite{H15_ghadi2024perf}.
%%%%%%%%%%%%%%%%%%%%%%%%%%%%%%%%%%%%%%%%%%%%%%%%%%%%%%%%%%%%%%%%%%%%%%%%%%%%%%%%
%\vspace{-3mm}

\begin{figure}
  \centering
  \subfigure[]{\includegraphics[width = \linewidth]{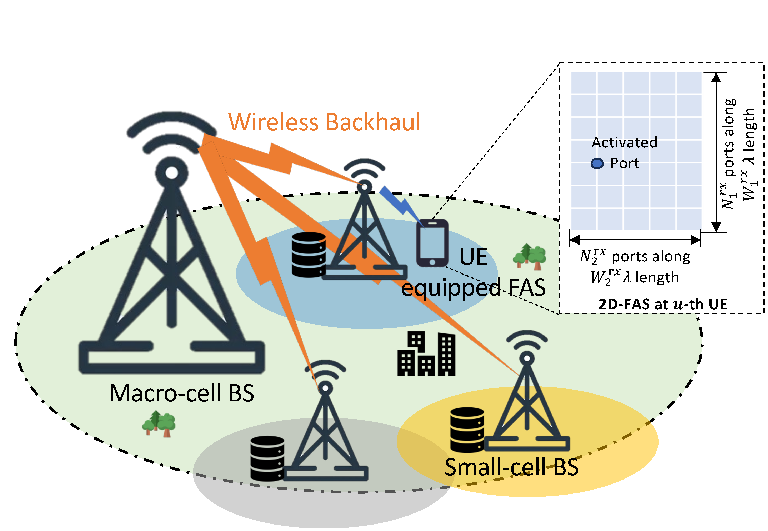}\label{Fig_CCN}}\vspace{-2mm}\\
  \subfigure[]{\includegraphics[width = \linewidth]{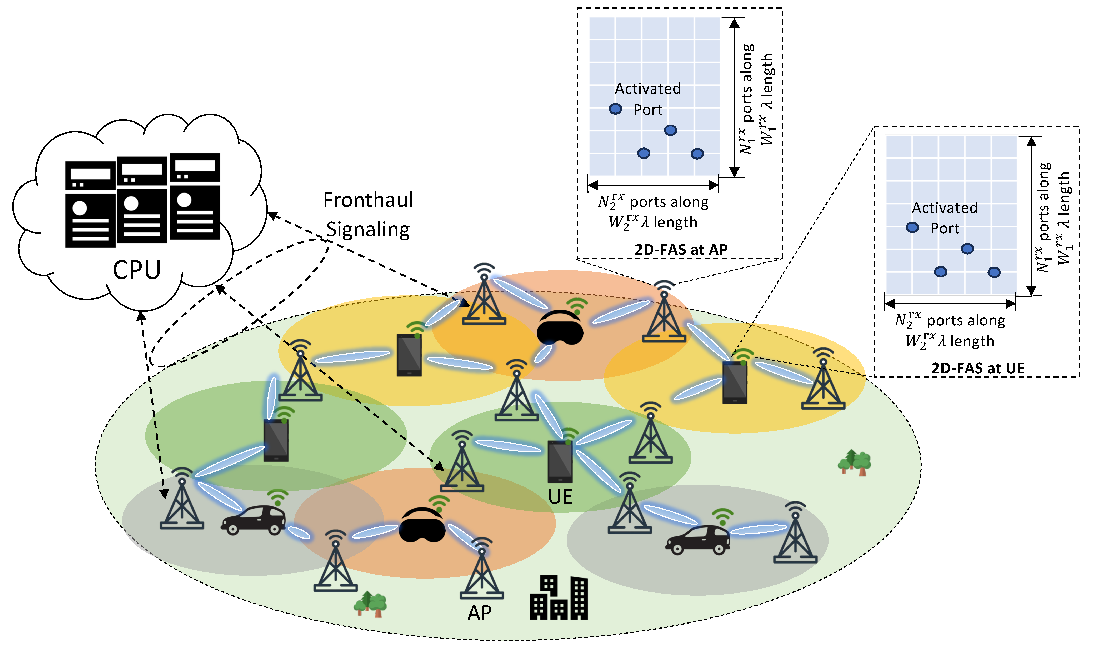}\label{Fig_cellfree}}\vspace{-2mm}
  \caption{Illustration of {\color{blue}(a) Case 3A}: FAS in a CCN {\color{blue}and (b) FAS-assisted cell-free network.}}\vspace{-2mm}
\end{figure}

\subsection{Case 3: HetNets} \label{subsec:HetNet}
\subsubsection{Case 3A: Content-centric FAS HetNets}\label{subsec:CCN}
Recent advancements have further introduced FAS in HetNet scenarios, particularly within the content-centric networks (CCNs)~\cite{F3_rostami2024cache}. In CCNs, frequently accessed content is cached at small-cell BSs (SBSs) and/or edge nodes~\cite{F2_bastug2014living}. This mechanism effectively reduces end-to-end latency and alleviates backhaul load, which is essential for future 6G networks that demand ultra-low latency and high throughput with minimal infrastructural overhead. As shown in Fig.~\ref{Fig_CCN}, content caching in CCN-enabled HetNets ensures that data is closer to end-users, minimizing redundant core network transmissions.

{\color{blue}The adaptability feature of FAS is especially beneficial in content-centric HetNets, where establishing stable and high-capacity links between SBSs and UEs is vital for efficient content delivery. FAS facilitates real-time optimization of channel conditions for UEs, effectively mitigating the detrimental impacts of fading and interference that frequently diminish content delivery performance associated with TAS. Therefore, the integration of FAS into CCN-enabled HetNets signifies a transformational advancement in next-generation networks, offering several significant benefits: (i) it enables adaptive link establishment, thereby allowing UEs to select the optimal antenna positions to enhance received signal strength during content retrieval; (ii) it improves spectral efficiency by dynamically adjusting the antenna configuration in response to network conditions, leading to more efficient resource utilization; and (iii) it enhances the reliability and robustness of wireless communication in 6G networks, particularly in high-mobility environments such as vehicular networks and IoT, where consistent connectivity and performance are of paramount importance.}

\subsubsection{\color{blue}Case 3B: FAS-assisted Cell-free Networks}
{\color{blue} As shwon in Fig.~\ref{Fig_cellfree}, the FAS-assisted cell-free network integrates the distributed intelligence of cell-free architectures with the spatial adaptability of FAS, thereby facilitating real-time optimization of signal propagation. The critical components of this integrated network consist of a central processing unit (CPU), FAS-equipped access points (APs), and FAS-equipped UEs. The CPU utilizes real-time CSI to capture optimal FAS ports. FAS equipped at APs is capable of dynamically adjusting their position, orientation, or radiation pattern to exploit spatial diversity for SE enhancement~\cite{shi2025cellfree}. FAS at each UE, on the other hand, can dynamically identify the port(s) with optimal AP clusters or propagation paths, mitigating the interference and enhancing uplink/downlink reliability. In particular, when each UE is served by its nearest adjacent AP, the network is characterized as a cell-free FAMA, as investigated in~\cite{han2025cellfree}. In the cell-free FAMA network, the FPAs at AP utilize maximum ratio transmission (MRT) to beam signals to their associated UE, while each UE relies on FAS to overcome interference. Analytical results demonstrate that the presence of FAS at each UE can enhance SE and substantially alleviate the CSI overhead burden on APs.}
%%%%%%%%%%%%%%%%%%%%%%%%%%%%%%%%%%%%%%%%%%%%%%%%%%%%%%%%%%%%%%%%%%%%%%%%%%%%%%%%
%\vspace{-3mm}

\subsection{Conclusion of Application Scenarios}
In this section, typical application scenarios are categorized into three types: homogeneous single-user, homogeneous multi-user, and heterogeneous networks with FAS. 

The majority of the state-of-art FAS scenarios are based on HomoNet, which primarily employs macro-cell deployment. We further delineate the HomoNet FAS scenario into two categories: single-user and multi-user HomoNet scenarios. The single-user HomoNet includes basic FAS, RIS-aided FAS, and FAS-aided IM. Basic FAS is designed to enhance the performance and capacity of the network. In the RIS-aided FAS network, FAS effectively addresses the double-path loss of RIS. Furthermore, FAS-aided IM contributes to improving spectral efficiency. In the context of the multi-user HomoNet with FAS, three typical application scenarios emerge: MU-MIMO-FAS, FAMA, and FAS-aided ISAC. MU-MIMO-FAS is capable of enhancing both robustness and capacity; FAMA facilitates massive connectivity; and FAS-aided ISAC not only improves spectral efficiency but also offers adaptive performance tuning.
In HetNet, FAS has recently been integrated into the content-centric HetNets {\color{blue} or cell-free networks}, enabling adaptive link establishment, enhancing spectral efficiency, and improving the reliability and robustness of the networks.

Additionally, AI technologies play a critical role within various FAS network scenarios~\cite{H18_wang2024AI}. The key challenge addressed by deep learning approaches in FAS networks pertains to the port selection problem~\cite{G3_chai2022SISO-FAS-PS, H9_Waqar2024Opportunistic, H20_zou2024online}. Deep learning methodologies can effectively tackle the high dynamics associated with channel adaptation in the port selection problem, yielding substantial improvements in spectral efficiency, connectivity, and flexibility within FAS-aided networks.
%%%%%%%%%%%%%%%%%%%%%%%%%%%%%%%%%%%%%%%%%%%%%%%%%%%%%%%%%%%%%%%%%%%%%%%%%%%%%%%%%%%%%%%%%%%%%%%%%%%%%%%%%%%%%%%%%%%%%%%%%%%%%%%%%%%%%%%%%%%%%%%%%%%%%%%%%%%%%%%%
%\vspace{-3mm}

\section{Fundamentals of FAS}\label{sec:fundamental}
This section presents an overview of the fundamentals of FAS, including its channel model, channel estimation, FAS, and FAMA. As FAS constitutes an emerging technology, channel modeling serves as a crucial aspect for assessing its performance. {\color{blue}Moreover, the foundational techniques developed for FAS-aided systems predominantly concentrate on either single-user FAS or multi-user FAMA. We reviews the transmission model and the performance of FAS and FAMA in this section.}
%%%%%%%%%%%%%%%%%%%%%%%%%%%%%%%%%%%%%%%%%%%%%%%%%%%%%%%%%%%%%%%%%%%%%%%%%%%%%%%%
%\vspace{-3mm}

\subsection{Channel Model} \label{subsec:chan} % Hao
We present channel models for both rich scattering and finite scattering environments, focusing on two FAS configurations: 1D-FAS and 2D-FAS. % These models effectively encapsulate the spatial correlation among FAS ports, as well as the performance characteristics specific to each antenna structure and propagation environment. 
Without loss of generality, we assume a point-to-point channel where the transmitter utilizes a single FPA and the receiver employs a FAS.

\subsubsection{Rich-Scattering Environment}
For the 1D-FAS scenario, the receiver has a 1D-FAS with $N$ ports, where the channel vector $\boldsymbol{h} = [ h_1, \dots, h_N ]^T$ captures the gains from the transmitter to FAS ports, with $h_n \sim {\cal CN}(0, \sigma^2)$. Due to the proximity of the ports, $\boldsymbol{h}$ exhibits strong spatial correlation, characterized by the covariance matrix $\boldsymbol{J} = \sigma^2 \boldsymbol{\varSigma}$. Following the Jake's model~\cite{Xu1_Jakes_1974}, the $(n,m)$-th element of $\boldsymbol{\varSigma}$ is given by
\begin{equation}\label{mn}
	(\boldsymbol{\varSigma})_{n,m} = \frac{1}{\sigma^2} {\rm Cov} \left[ h_n, h_{m} \right] = J_0 \left( \frac{2 \pi (n-m)}{N - 1} W \right),
\end{equation}
where $J_0 (\cdot)$ denotes the zero-order Bessel function of the first kind. For performance analysis, the channel model $\boldsymbol{h}$ must capture spatial correlation properties while remaining analytically tractable. {\color{blue}Previous works~\cite{I21_wong2022FAS} and~\cite{H4_wong2022FAMA} proposed a simplified FAS channel model. This model can simplify FAS performance analysis, but may lead to overoptimistic estimation. The channel coefficients of this simplified model for the $n$-th port can be expressed as}
\begin{multline}\label{model1}
h_n= \sigma \left( \sqrt{1 - \mu_n^2} x_n + \mu_n x_0 \right)\\
 + j \sigma \left( \sqrt{1 - \mu_n^2} y_n + \mu_n y_0 \right), ~ {\text {for}}~ n = 1, \dots, N,
\end{multline}
where $x_0, \dots, x_N, y_0, \dots, y_N$ follow independent and identically distributed (i.i.d.) ${\cal N}(0, \frac{1}{2})$, and $\mu_n$ is simplified as
\begin{equation}\label{mu_n}
\mu_n = J_0 \left( \frac{2 \pi (n-1)}{N - 1} W \right).
\end{equation}
While this model relies on the first port as a reference, it may potentially overlook interdependencies among other ports, as noted in (\ref{mn}), which may lead to overly optimistic performance evaluations.

Several studies have enhanced the channel model (\ref{model1}). In particular, Wong {\em et al.} introduced an approach that that replaces individual correlation parameters with a common parameter, $\mu$, defined by~\cite{wong2022closed}
\begin{equation}\label{mu}
\mu = \sqrt{2} \sqrt{_1F_2 \left( \frac{1}{2}; 1, \frac{3}{2}; -\pi^2 W^2 \right) - \frac{J_1(2\pi W)}{2\pi W}},
\end{equation}
where $_1F_2 ( \cdot; \cdot; \cdot)$ is the generalized hypergeometric function and $J_1 (\cdot)$ is the first-order Bessel function of the first kind. This approach establishes mutual correlations among all ports within a FAS, allowing any port to serve as a reference. Building upon this model,~\cite{wong2022closed} conducted a thorough analysis of both outage probability and multiplexing gain, while~\cite{chai2022performance} explored optimal port selection strategies. However, the channel model provided by~\cite{wong2022closed} still inadequately reflects the correlation among antenna ports as characterized by (\ref{mn}).

To improve accuracy,~\cite{khammassi2022new} introduced an innovative approach in which each channel coefficient is represented as a weighted linear combination of $N$ i.i.d.~complex Gaussian random variables. Specifically, let $\boldsymbol{U}\!\boldsymbol{\varLambda} \boldsymbol{U}^H$ denote the eigen-decomposition of $\boldsymbol{\varSigma}$. The channel can be expressed as
\begin{equation}\label{h_jk}
	\boldsymbol{h} = \sigma \boldsymbol{U} \boldsymbol{\varLambda}^{\frac{1}{2}} \boldsymbol{g},
\end{equation}
where $\boldsymbol{g} = [ g_1, \dots, g_N ]^T \sim{\cal CN}(\boldsymbol{0}, \boldsymbol{I}_N)$. Accordingly, the $n$-th element of $\boldsymbol{h}$ can be expressed as 
\begin{align}\label{h_jkn}
	h_n & = \sigma \sum_{m = 1}^N \sqrt{\lambda_m} u_{n,m} g_n \nonumber\\
	& = \sigma \sum_{m = 1}^N \sqrt{\lambda_m} u_{n,m} \left( a_n + i b_n \right),
\end{align}
where $u_{n,m}$ is the $(n,m)$-th element of $\boldsymbol{U}$, and $a_n$ and $b_n$ are random variables distributed as ${\cal N}(0, \frac{1}{2})$. It is evident that $\boldsymbol{h}$ designed in (\ref{h_jk}) or (\ref{h_jkn}) adheres to the specified distribution, i.e., $\boldsymbol{h} \sim {\cal CN}(\boldsymbol{0}, \sigma^2 \boldsymbol{\varLambda})$. Nevertheless, while this ``perfect'' model offers theoretical precision, it is constrained by complex nested integrals as documented in~\cite{hao2023on}. To simplify,~\cite{khammassi2022new} and~\cite{hao2023on} proposed a simplification model. The covariance matrix $\boldsymbol{\varSigma}$ of $\boldsymbol{h}$ exhibits a Hermitian Toeplitz structure, where the eigenvalue concentration allows approximation using ${\color{blue}N_e} \ll N$ eigenvalues.
% In particular, $h_n$ in (\ref{h_jkn}) can be approximated by
% \begin{align}\label{model3}
% 	{\hat h}_n = \sigma \sum_{m = 1}^M \sqrt{\lambda_m} u_{n,m} \left( a_n + i b_n \right),
% \end{align}
Using the model,~\cite{khammassi2022new,hao2023on} analyzed the outage probability for single- and two-user FASs, respectively.

For 2D-FAS, the receiver employs a 2D-FAS of size $W_1\lambda \times W_2\lambda$, which has a uniform grid structure with $N = N_1 \times N_2$ ports. Ports are indexed from left to right and bottom to top along, with index assignment given by
\begin{equation}\label{k_n1n2}
	n = (n_2 - 1)N_1 + n_1. % _{(n_1, n_2)}
\end{equation}
The element $(\boldsymbol{\varSigma})_{n,m}$ describing the correlation between the ports $(n_1, n_2) \to n$ and $(m_1, m_2) \to m$ in the correlation matrix $\boldsymbol{J} = \sigma^2\boldsymbol{\varSigma} \in {\mathbb C}^{N \times N}$ is given by
\begin{equation}\label{J_tx_ele}
(\!\boldsymbol{\varSigma})_{n,m}= j_0 \!\!\left(\!\! 2\pi \! \sqrt{\!\left( \frac{n_1 - m_1 }{N_1-1}W_1 \right)^2 \!+\! \left( \frac{n_2 - m_2}{N_2-1} W_2  \right)^2 } \right)\!\!,
\end{equation}
where $j_0 (\cdot)$ is the spherical Bessel function of the first kind. Similarly, with the eigen-decomposition $\boldsymbol{\varSigma} = \boldsymbol{U}\boldsymbol{\varLambda}\boldsymbol{U}^H$, the complex channel vector can be modeled as in~\eqref{h_jk} or~\eqref{h_jkn}~\cite{G5_new2023SISO-FAS}.
% \begin{equation}\label{H}
% 	\boldsymbol{h} = \boldsymbol{U} \boldsymbol{\varLambda}^{\frac{1}{2}} \boldsymbol{g},
% \end{equation}
% where $\boldsymbol{g} \in {\mathbb C}^{N \times 1} $ follows ${\cal CN}(\boldsymbol {0}, \boldsymbol{I}_{N})$, $\boldsymbol{U}$ is an $N \times N$ matrix with columns being the eigenvectors of $\boldsymbol{J}$, and $\boldsymbol{\varLambda} = \mathrm{diag}(\lambda_1, \ldots, \lambda_{N})$ is the matrix with diagonal entries being the corresponding eigenvalues.

In~\cite{G10_new2024MIMO-FAS}, the channel model is extended to the case where both the transmitter and the receiver use a FAS, and multiple ports can be activated. In this case, SISO-FAS is extended to MIMO-FAS, and the complex channel becomes
\begin{equation} \label{eq-MIMO-channel}
    \boldsymbol{H}= \boldsymbol{U}_{\text{rx}} \sqrt{\boldsymbol{\varLambda}_{\text{rx}}}\boldsymbol{G} \sqrt{\boldsymbol{\varLambda}_{\text{tx}}^H} \boldsymbol{U}_{\text{tx}}^H,
\end{equation}%
where $\boldsymbol{G} \in \mathbb{C}^{N_{\text{\text{rx}}} \times  N_{\text{tx}}}$ with each entry being i.i.d.~and following the distribution $\mathcal{CN}(0, 1)$, $\boldsymbol{U}_\text{s}$ is an $N_\text{s} \times N_\text{s}$ matrix whose columns are the eigenvectors of $\boldsymbol{J}_\text{s}$ and $\boldsymbol{\varLambda}_\text{s}= \mathrm{diag}(\lambda_1^\text{s}, \ldots, \lambda^\text{s}_{N_\text{s}})$ is an $N_\text{s} \times N_\text{s}$ diagonal matrix whose diagonal entries are the corresponding eigenvalues, ${\rm s}\in \{{\rm tx},{\rm rx}\}$.

\subsubsection{Finite-Scattering Environment}
In a finite-scattering environment such as millimeter-wave communication systems, the planar-wave geometric model effectively characterizes the channel~\cite{akdeniz2014millimeter}. In this case, $\boldsymbol{h}$ is modeled as
\begin{equation}\label{h}
	\boldsymbol{h} = \sqrt{N} \sum_{l = 1}^{L} \gamma_l \boldsymbol{a}_l,
\end{equation}
where $L$ is the number of propagation paths and $\gamma_l$ is the complex channel gain of the $l$-th path, $\boldsymbol{a}_l$ is the steering vector at the receiver. This steering vector for 1D-FAS is given by
\begin{equation}\label{a_RT}
		\boldsymbol{a}_l \!=\! \frac{1}{\sqrt{N}} \left[ 1, e^{- j \frac{2 \pi}{\lambda} \Delta \cos \phi_l}, \dots, e^{- j \frac{2 \pi}{\lambda} (N-1) \Delta \cos \phi_l} \right]^{T},
\end{equation}
where $\phi_l \in [0, \pi]$ is the angle-of-arrival (AoA) of the $l$-th propagation path and $\Delta = {W\lambda}/{(N - 1)}$ is the distance between any two adjacent ports.

In 2D-FAS, the FAS surface lies on the x-y plane. 
{\color{blue}The $(1,1)$-th port is selected as the reference port, and its center is placed at the origin of the coordinate system with three-dimensional (3D) coordinate vector $[0,0,0]$.}
The vector of the $(n_1,n_2)$-th port within the three-dimensional (3D) coordinate is therefore $\left[ (n_1 - 1)\Delta_1, (n_2 - 1)\Delta_2, 0 \right]$, where $\Delta_1 = {W_1\lambda}/{(N_1-1)}$ and $\Delta_2 = {W_2\lambda}/{(N_2-1)}$. Let the elevation and azimuth AoAs for the $l$-th propagation path be denoted by $\theta_l \in [0, \pi]$ and $\phi_l \in [0, \pi]$, respectively. The propagation difference of the $l$-th path between ports $(1,1)$ and $(n_1,n_2)$ is then given by
\begin{equation}\label{diff_prop}
	s_l (n_1, n_2) \!=\! (n_1 \!-\! 1)\Delta_1 \!\sin \theta_l \cos \phi_l \!+\! (n_2 \!-\! 1)\Delta_2 \! \cos \theta_l,
\end{equation}
resulting in phase difference $\frac{2 \pi}{\lambda} s_l (n_1, n_2)$. The steering vector for 2D-FAS is defined as
\begin{equation}\label{a_T_2D}
	\boldsymbol{a}_l = \frac{1}{\sqrt{N}} \left[ 1, e^{- j \frac{2 \pi}{\lambda} s_l (1, 2)}, \dots, e^{- j \frac{2 \pi}{\lambda} s_l (N_1, N_2)} \right]^{T}.
\end{equation}
% The channel matrix of 2D-FAS is then modeled as in~\eqref{h} with the updated steering vector.
% \begin{align}\label{H_mmwave}
% 	\boldsymbol{H} = \sqrt{N} \sum_{l = 1}^{L} \gamma_l  \boldsymbol{a}_l^H.
% \end{align}
\subsubsection{\color{blue}Conclusion of channel models}
{\color{blue}
This section presents channel models for both rich and finite scattering environments. The channel model in the rich scattering environment is based on the statistical model. It can simplify the performance evaluation and is accurate in the rich scattering environment for the sub-6GHz band. In the mmWave band, the scattering environment is expected to be more sparse, and a planar-wave geometric model is used to describe the FAS channel model in this section.
}
%%%%%%%%%%%%%%%%%%%%%%%%%%%%%%%%%%%%%%%%%%%%%%%%%%%%%%%%%%%%%%%%%%%%%%%%%%%%%%%%

\subsection{Channel Estimation} % Hao
To achieve best performance, reliable CSI for all ports is essential. Nevertheless, estimating CSI for all ports can lead to substantial hardware switching and system overhead. This subsection outlines the channel estimation process for FAS in rich-scattering and finite-scattering environments.

\subsubsection{Rich-Scattering Environment}
Several papers have investigated channel estimation in the rich-scattering environment. For instance,~\cite{skouroumounis2022fluid} and~\cite{skouroumounis2022large} investigated channel estimation and outage performance of large-scale cellular networks using circular multi-fluid antenna arrays. They proposed the skipped-enabled linear minimum mean square error (LMMSE)-based channel estimation (SeCE) technique, which reduces signaling overhead by estimating CSI for a select subset of ports, though it results in lower observed SINR and involves a trade-off between overhead reduction and SINR degradation.

Recent studies have highlighted the role of machine learning techniques in channel estimation for FAS. For instance,~\cite{ji2024correlation} showed how machine learning could recover full CSI from a limited number of ports, introducing specialized sub-networks for different conditions. In addition, they proposed a hard selection method to adjust the number of estimated ports based on correlation levels.
In~\cite{zhang2024learning}, the authors investigated spatial extrapolation using an asymmetric graph masked autoencoder (AGMAE) framework to address the complexities of high-resolution FAS, combining graph-based processing with an architecture that enhances generalization capability.

\subsubsection{Finite-Scattering Environment}
The channel estimation problem in finite-scattering environments has been studied in several works~\cite{xu2024channel, wang2023estimation, ma2023compressed, xiao2023channel, zhang2023successive}. Here we focuse on a typical multi-user uplink system described in~\cite{xu2024channel}. In this context, each user employs a 1D-FAS, while the BS has $\color{blue}N_{\rm BS}$ FPAs. Let $\boldsymbol{h}_{u,n} \in {\cal C}^{M \times 1}$ denote the channel vector from the $n$-th port of user~$u$ to the BS. By stacking $\boldsymbol{h}_{u,n}$ for all $n \in \{1, \dots, N\}$, we obtain the channel matrix between all ports of user~$u$ and the BS as
\begin{equation}\label{H_u}
\boldsymbol{H}_u = [\boldsymbol{h}_{u,n}, \dots, \boldsymbol{h}_{u,N}].
\end{equation}
We introduce various schemes to estimate $\boldsymbol{H}_u$ below.

% \begin{figure}[]
% 	\centering
% 	\includegraphics[scale=0.35]{Figure/SecIII/FigIII.B.1_system_channel_est.pdf}\vspace{-3mm}
% 	\caption{Channel estimation for a multi-user uplink system, where each user is equipped with a linear FAS, while the BS has multiple fixed-position antennas. }\label{system_channel_est}\vspace{-5mm}
% \end{figure}

\paragraph{Least Square (LS) Scheme}
A basic approach to estimate $\boldsymbol{H}_u$ involves all users sequentially transmitting orthogonal pilot sequences across $N$ ports, each repeated $T$ times. The LS estimate of $\boldsymbol{h}_{u,n}$ can be obtained by right-multiplying the normalized signal matrix with the corresponding pilot sequence, given by
\begin{align}\label{y_nt_qu}
{\hat {\boldsymbol{h}}}_{u,n}^{\text {LS}} = \boldsymbol{h}_{u,n} + {\hat {\boldsymbol{z}}}_{u,n},
\end{align}
where ${\hat {\boldsymbol{z}}}_{u,n}$ is the estimation error caused by the additive noise. By stacking (\ref{y_nt_qu}) for all $n$, the LS estimate of $\boldsymbol{H}_u$ is
\begin{align}\label{Gu_LS}
{\hat {\boldsymbol{H}}}_{u, {\text {LS}}} = \boldsymbol{H}_u + {\hat {\boldsymbol{Z}}}_u.
\end{align}
The accuracy of LS largely depends on the number of times a pilot sequence is repeated, i.e., $T$, and the transmit power. However, given that the number of ports $N$ is typically large, LS method suffers from high hardware complexity and pilot overhead, making it impractical for real-world applications.

\paragraph{Low-Sample-Size Sparse Channel Reconstruction (L3SCR) Method}
In a finite-scattering environment, the channel exhibits sparsity. In sparse channel environments, effective estimation can occur by limiting measurement to $K \ll N$ preset estimating locations (ELs). The channel vector from the $k$-th EL of user $u$ is $\boldsymbol{g}_{u,k} \in {\mathbb C}^{M \times 1}$, and denote $\boldsymbol{G}_u = [\boldsymbol{g}_{u,n}, \dots, \boldsymbol{g}_{u,K}]  \in {\mathbb C}^{M \times K}$. Using the planar-wave geometric channel model, $\boldsymbol{G}_u$ can be expressed as
\begin{equation}\label{G}
	\boldsymbol{G}_u = \sqrt{MK} \sum_{l = 1}^{L_u} \gamma_{u,l} \boldsymbol{a}_{u,{\text R}} (\phi_{u,l}) \boldsymbol{a}_{u,{\text T}}^H (\theta_{u,l}),
\end{equation}
where $L_u$ is the number of propagation paths between user $u$ and the BS, $\gamma_{u,l}$ is the complex channel gain of the $l$-th path, and $\phi_{u, l}, \theta_{u, l} \in [0, \pi]$ are the corresponding AoA and AoD. In addition, $\boldsymbol{a}_{u,{\text R}} (\phi_{u, l})$ and $\boldsymbol{a}_{u,{\text T}} (\theta_{u, l})$ are respectively the steering vectors at the receiver and transmitter sides. In~\cite{xu2024channel}, a method called L3SCR is proposed to estimate the sparse channel parameters from $\boldsymbol{G}_u$ and reconstruct $\boldsymbol{H}_u$.
{\color{blue}This method consists of three main steps: First, the number of propagation paths and AoAs are estimated using a combination of the DFT-based method and an angle rotation operation. Then, the AoDs and channel gains of different paths are estimated using matched filtering techniques. Finally, based on the estimated parameters and the planar-wave geometric model, the complete channel matrix $\boldsymbol{H}_u$ is reconstructed.}

\paragraph{Orthogonal Matching Pursuit (OMP) Algorithm}
The accuracy of the L3SCR method depends significantly on the number of BS antennas, $\color{blue}N_{\rm BS}$. When $\color{blue}N_{\rm BS}$ is sufficiently large, the estimates of the number of paths and AoAs are accurate. Otherwise, the accuracy of these estimates may be compromised. To mitigate this issue, the OMP algorithm can be utilized to enhance estimation accuracy, albeit at the expense of increased computational complexity. OMP jointly estimates the AoD and AoA in an iterative manner.
{\color{blue}The OMP algorithm works in five steps: First, similar to the L3SCR scheme, all users' FAS transmit orthogonal pilot sequences at $K$ ELs, allowing for the least squares (LS) estimation of $\boldsymbol{G}_u$. Second, a set of quantized angle grids is selected, based on which the sensing matrix is constructed. Third, the column of the sensing matrix that exhibits the strongest correlation with the residual vector is identified. The index of this column corresponds to a pair of quantized angles, from which an AoD/AoA pair can be estimated. Fourth, the channel gain associated with the identified AoD/AoA pair is estimated by solving a least squares problem. In the fifth step, the residual vector is updated by subtracting the contributions of the chosen column vectors.}
Iteratively carry out the third to the fifth steps until the difference between consecutive residual vectors falls below a predefined threshold. Once the sparse parameters have been accurately estimated, the final step is to reconstruct the complete channel matrix using the approach employed in the L3SCR scheme.

% \begin{figure}
% 	\centering
% 	\includegraphics[width=\linewidth]{Figure/SecIII/FigIII.B.2_NMSE_VS_KM.eps}
% 	\caption{NMSE versus the number of ELs $K$ with SNR = $10$ dB and $T = 1$.}\label{NMSE_VS_KM}
% \end{figure}
\begin{figure}
	\centering
	\includegraphics[width=\linewidth]{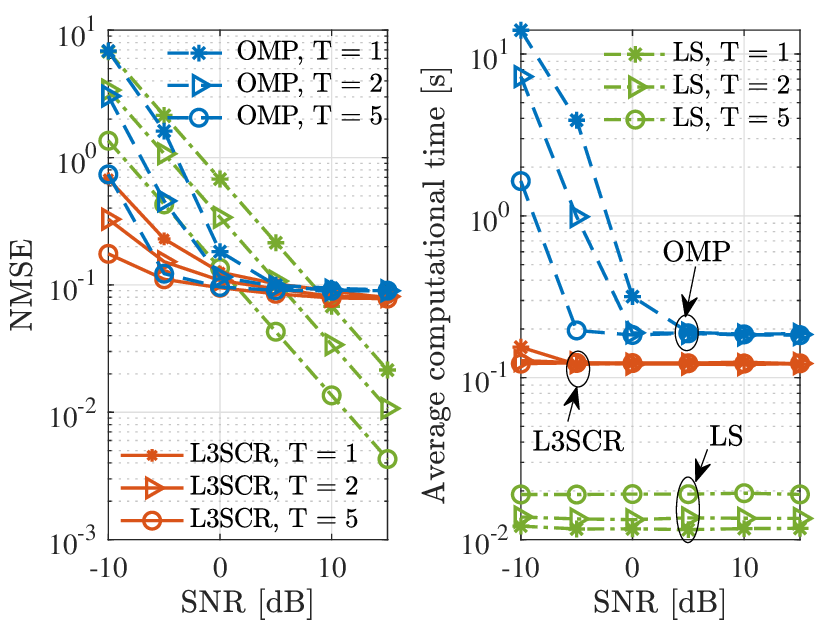}%\vspace{-3mm}
	\caption{NMSE and computational time with ${\color{blue}N_{\rm BS}} = 64$ and $K = 10$.}\label{NMSE_time_VS_PT}\vspace{-2mm}
\end{figure}

% Fig.~\ref{NMSE_VS_KM} and 
Fig.~\ref{NMSE_time_VS_PT} illustrate the normalized mean square error (NMSE) and the average computational time for one channel realization, comparing different schemes across various parameters.
% First, when $\color{blue}N_{\rm BS}$ is small, the OMP algorithm outperforms the L3SCR scheme in terms of NMSE, while the situation reverses when $\color{blue}N_{\rm BS}$ is large, as previously discussed.
In many scenarios, the LS method achieves a lower NMSE compared to both the L3SCR and OMP schemes. However, it is crucial to note that the LS method incurs significantly higher hardware switching and pilot overhead, as it requires the antennas of all users to switch and transmit pilot sequences across all $N$ ports. This increased overhead can lead to a reduction in spectral efficiency. Additionally, it is evident that the OMP algorithm requires much more computational time than the L3SCR scheme. This is due to the fact that the OMP algorithm involves matrix inversions when solving LS problems, which is computationally intensive.

\paragraph{Other Schemes}
In the previous two paragraphs, we show that in the finite-scattering environment, the geometric model can be adopted to characterize the channel of an FAS-assisted uplink system. In this case, standard tools can be used to estimate the sparse parameters at some preset ELs, based on which the CSI at all $N$ ports can be reconstructed. In addition to the L3SCR and OMP methods, several other schemes can also be utilized for channel estimation in FAS-assisted systems. These include methods such as the multiple signal classification (MUSIC) technique~\cite{guo2017millimeter}, the estimation of signal parameters via rotational invariance technique (ESPRIT)~\cite{roy1989esprit}, the unitary ESPRIT algorithm~\cite{haardt1995unitary}, and the space-alternating generalized expectation-maximization (SAGE) scheme~\cite{fleury1996wideband}, among others. These methods primarily differ in terms of the accuracy they provide in estimation and the computational complexity required for their implementation.
%%%%%%%%%%%%%%%%%%%%%%%%%%%%%%%%%%%%%%%%%%%%%%%%%%%%%%%%%%%%%%%%%%%%%%%%%%%%%%%%

\subsection{Single-user FAS and its Variants} % Xinghao
\subsubsection{Transmission Model}
For a point-to-point FAS where the transmitter uses a fixed antenna, but the receiver has a fluid antenna with one activated port (i.e., SISO-FAS), the received signal at the $n$-th port of fluid antenna is given by
\begin{equation}\label{eq:siso-fas}
  r_n = h_n s +\eta_n,
\end{equation}
where $h_n$ represents the complex channel coefficient at the $n$-th port, $s$ is the transmitted information-bearing symbol and $\eta_n \sim \mathcal{CN}(0,\sigma_\eta^2)$ is the additive white Gaussian noise (AWGN). To achieve optimal performance, FAS activates the port corresponding to the strongest channel, i.e., 
\begin{equation}\label{eq-SISO-channel}
h_\mathrm{FAS} = \max_n \{|h_n|\}.
\end{equation}%
After selecting the optimal port, the rate of the SISO-FAS can be expressed as
\begin{equation}
\label{eq-SISO-rate}
    R = \log\left(1+\frac{\sigma_s^2}{\sigma_\eta^2}\left|h_\mathrm{FAS}\right|^2\right),
\end{equation}%
in which $\sigma_s^2 = \mathbb{E}\{|s|^2\}$. The outage probability of the SISO-FAS can be given by
\begin{equation}
\label{eq-SISO-Pout}
    p_\mathrm{SISO-FAS} = {\rm Prob}\left(\frac{\sigma_s^2}{\sigma_\eta^2}\left|h_\mathrm{FAS}\right|^2<\gamma\right),
\end{equation}%
where $\gamma$ represents the SNR threshold. The outage probability of the SISO-FAS is derived in~\cite{G5_new2023SISO-FAS}. At high SNR the outage probability of SISO-FAS is given by~\cite[Theorem 2]{G5_new2023SISO-FAS}
\begin{equation}
  p_\mathrm{SISO-FAS} = \frac{1}{\det(\boldsymbol{J})}{\left(\frac{\gamma}{\Gamma}\right)}^N +o\left(\Gamma^{-N}\right),
\end{equation}
where $\Gamma = \sigma_s^2/\sigma_\eta^2$ is the transmit SNR.

As illustrated in \figref{fig-su-homo}(a), fluid antenna can be adopted at both transmitter and receiver sides, and multiple ports can be activated. In this MIMO-FAS case, the channel of MIMO-FAS is given by~\eqref{eq-MIMO-channel}. The activation port matrix is denoted as $\boldsymbol{A}_\text{s} = [\boldsymbol{a}_1^\text{s}, \dots,  \boldsymbol{a}_{n_\text{s}}^\text{s}]$, where $\boldsymbol{a}_l^\text{s}$ is the $N_\text{s}$-dimensional standard basis vector for $l \in \{1,2, \ldots, n_\text{s}\}$. The beamforming matrix is denoted as ${\boldsymbol{W}}_{\text{s}}$ with the constraint ${\| \boldsymbol{W}_\text{s} \|}_2= 1$. Then, the received signal of MIMO-FAS can be written as 
\begin{align}\label{eq-MIMO-signal}
  \tilde{\boldsymbol{Y}} & = \boldsymbol{W}_{\text{rx}}\boldsymbol{A}_{\text{rx}}\boldsymbol{Y} \nonumber \\ &= \boldsymbol{W}_{\text{rx}}\boldsymbol{A}_{\text{\text{rx}}}\boldsymbol{H}\boldsymbol{A}_{\text{tx}}\boldsymbol{W}_{\text{tx}}\boldsymbol{s} + \boldsymbol{W}_{\text{rx}}\boldsymbol{A}_{\text{rx}}\boldsymbol{\eta}\nonumber \\ &\triangleq \tilde{\boldsymbol{H}}\boldsymbol{x} + \tilde{\boldsymbol{\eta}},
\end{align}
where $\boldsymbol{s}$ is the transmitted signal and $\boldsymbol{\eta}$ is the AWGN. Denote $\bar{\boldsymbol{H}}= \boldsymbol{A}_{\text{rx}}\boldsymbol{H}\boldsymbol{A}_{\text{tx}}$ and $\boldsymbol{K}= \boldsymbol{W}_{\text{tx}}\boldsymbol{P}\boldsymbol{W}_{\text{tx}}^H$, where $\boldsymbol{K}$ is the input covariance and $\boldsymbol{P}= \mathbb{E}[\boldsymbol{s} \boldsymbol{s}^H]$ is the power allocation matrix. Then, the rate of MIMO-FAS can be expressed as
\begin{equation}
\label{eq-MIMO-rate}
R=\log \operatorname{det} \left(\boldsymbol{I} + \bar{\boldsymbol{H}} \boldsymbol{K} \bar{\boldsymbol{H}}^H\right),
\end{equation}%
where the value of $\operatorname{trace}(\boldsymbol{K})$ does not exceed the transmit SNR. The rate of MIMO-FAS can be maximized through joint optimization of port selection, transmit and receive beamforming, and power allocation. The solution can be obtained using an exhaustive search, singular value decomposition (SVD), and waterfilling for port selection, beamforming and power allocation, respectively, at the cost of high complexity. In~\cite{G10_new2024MIMO-FAS}, the joint optimization of the port activated matrices and beamforming matrices was addressed.

\begin{figure}
  \centering
  \includegraphics[width =\linewidth]{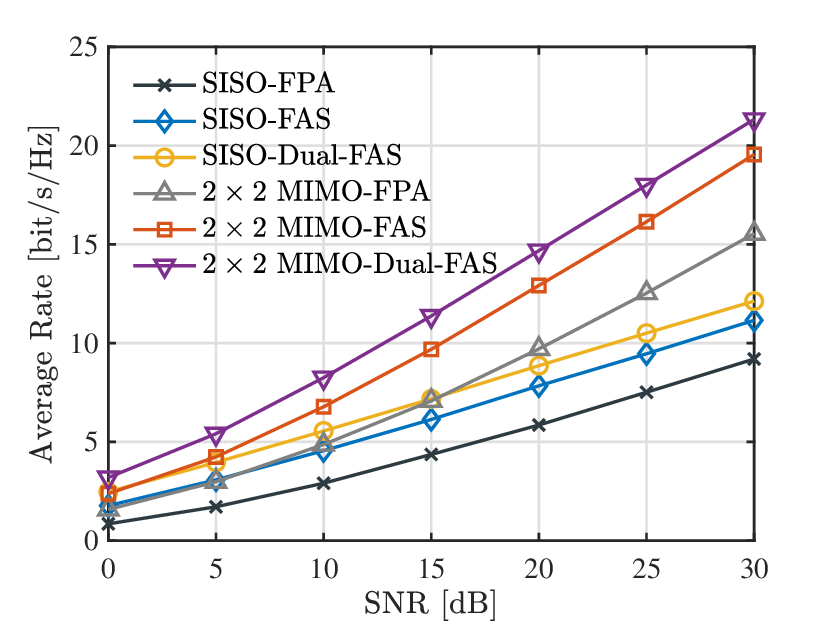}
  \caption{Channel capacity of SISO-FAS and MIMO-FAS against SNR for FAS with $W_1^\text{s} = W_2^\text{s} = 0.5 \lambda$ and $N_1^\text{s} = N_2^\text{s} = 4$ (${\rm s} \in \{{\rm tx},{\rm rx}\}$).}\label{Fig_FAS_Capacity}\vspace{-2mm}
\end{figure}

\subsubsection{Theoretical Performance}
\figref{Fig_FAS_Capacity} shows the rate of FAS compared to traditional FPA systems, where the ``FAS'' curves represent systems that employ a single fluid antenna at either the transmitter or receiver, and the ``Dual-FAS" curves represent systems where fluid antennas are employed at both the transmitter and receiver. Therefore, the SISO-Dual-FAS curve can be viewed as a special case of MIMO-FAS when $n_\text{s} = 1$. Also, the fluid antennas all have the same specifications, i.e., $W_1^\text{s} = W_2^\text{s} = 0.5 \lambda$ and $N_1^\text{s} = N_2^\text{s} = 4$, ${\rm s} \in \{{\rm tx},{\rm rx}\}$. It can be observed that with the assistance of FAS, the channel capacity can be significantly improved, and when fluid antennas are simultaneously employed at the transmitter and receiver, the channel capacity can be further enhanced.

\begin{figure}
  \centering
  \includegraphics[width =\linewidth]{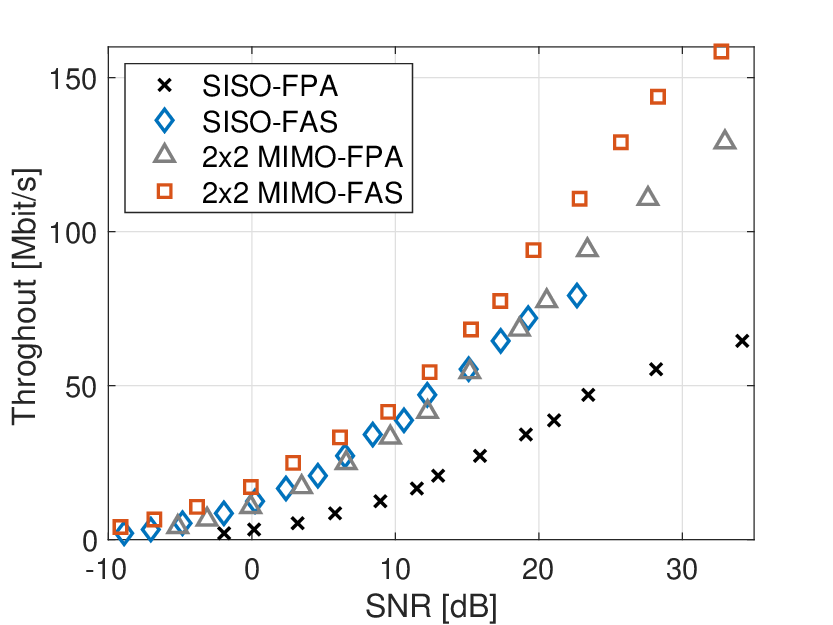}
  \caption{Throughput against SNR in link-level simulations of $20$ MHz bandwidth for FAS with $W_1^\text{tx} = W_2^\text{tx} = 0.5 \lambda$ and $N_1^\text{tx}= N_2^\text{tx} = 4$.}\label{Fig_FAS_LLS}\vspace{-2mm}
\end{figure}

\subsubsection{Link-level Simulation Results}
Recently,~\cite{Hong2025fluid} studied the use of FAS into the wideband 5G New Radio (NR) OFDM system and showed that FAS achieves significant gain in wideband systems. Results in Fig.~\ref{Fig_FAS_LLS} show the throughput of FAS with comparison to traditional FPA systems for different modulation and coding schemes. FAS is equipped at the receiver. The simulation bandwidth is set to be $20$ MHz, with the delay spread ${\rm DS} = 30$ ns and Doppler frequency $f_D = 30$ Hz. For FAS configuration, $N = 4 \times 4$ ports are distributed among FAS plane with the size of $W = 0.5\lambda \times 0.5\lambda$. It can be observed that FAS can improve the throughput in wideband channels. Specifically, SISO-FAS nearly double the throughput compared with FPA, and MIMO-FAS can further improve the throughput with a bit higher SNR.

\subsubsection{Variants of FAS}
The performance of FAS can be further enhanced with RIS. As seen in \figref{fig-su-homo}(b), in RIS-aided FAS, the received signal at the \(n\)-th port of the UE is written as
\begin{equation} \label{eq-RIS-signal}
y_n =\sigma_s\sum_{m=1}^M h_m g_{m, n} e^{-j \theta_m} s + \eta_n,
\end{equation}%
where $h_m \sim \mathcal{CN}(0, \sigma_h^2)$ denotes the channel coefficient from the BS to the \(m\)-th RIS element, \( g_{m,n} \sim \mathcal{CN}(0, \sigma_g^2) \) denotes the channel from the \(m\)-th RIS element to the \(k\)-th FAS port of the UE, $\theta_m$ denotes the reflecting phase of the $m$-th RIS element, {\color{blue}$m = 1,\dots,M_{\rm RIS}$, $M_{\rm RIS}$} is the number of RIS elements. By adjusting $\theta_m$ such that $\theta_m = -\arg (h_m g_{m,n})$, the channel gain can be maximized. As such, the UE will select and activate the port with the maximum combined channel gain, i.e.,
\begin{equation}
\label{eq-RIS-port}
    n^* = \arg\max_k \sum_{m=1}^{M_{\rm RIS}} |h_m| |g_{m, n}|.
\end{equation}
% \begin{figure}[t]
% \centerline{\includegraphics[width = 0.9\linewidth]{Figure/SecIII/FigIII.C.3_RIS_FAS_OutageProbability.eps}}\vspace{-3mm}
% \caption{Outage probability versus the number of ports $N$ for different number of RIS elements $M_{\rm RIS}$.}\label{fig-RIS-FAS-Pout}\vspace{-5mm}
% \end{figure}
% \figref{fig-RIS-FAS-Pout} provides the outage probability of the RIS-aided FAS against the number of ports $N$, and compares it with the scenario where the receiver employs FPA. The relevant parameters are set as $\sigma^2_s = 1$, $\sigma_\eta^2 = 0.01$, and the minimum rate threshold is $16$. As observed, increasing the number of ports significantly reduces the outage probability. Similarly, increasing the number of RIS elements provides additional improvements. These results demonstrate the synergistic relationship between FAS and RIS, which can be effectively leveraged to significantly enhance the performance.

When FAS is implemented only at the transmitter, combining with IM can effectively enhance spectral efficiency. As illustrated in \figref{fig-su-homo}(c), the transmitted signal vector $\boldsymbol{x} \in \mathbb{C}^{N_{tx} \times 1}$ in~\cite{G27_Zhu2024FA-IM} can be represented as
\begin{equation}
\label{eq-FA-IM-signal-t}
\boldsymbol{x}=\sum_{i=1}^{n_{tx}} s_i \boldsymbol{e}_{I_i},
\end{equation}%
where $s_i \in \chi$, $\chi$ is a $\color{blue}2^{Q_m}$-ary modulation constellation, and $\boldsymbol{e}_{I_i}$ denotes {\color{blue}the $I_i$-th column of the identity matrix $\boldsymbol{I}_{N_\text{tx}}$ of size $N_\text{tx}$}. The received signal vector $\boldsymbol{y} \in \mathbb{C}^{N_\text{rx} \times 1}$ is given by
\begin{equation}
\label{eq-FA-IM-signal-r}
\boldsymbol{y} = \boldsymbol{H}\boldsymbol{x} + \boldsymbol{\eta},
\end{equation}%
where $\boldsymbol{H}$ represents the complex channel, which is given by~\eqref{eq-MIMO-channel}, and $\boldsymbol{\eta}$ is the AWGN. Assuming the receiver has perfect knowledge of the CSI, the optimal maximum likelihood (ML) detector can be expressed as 
\begin{align}
\label{eq-FA-IM-ML}
    (\hat{\mathcal{I}},\hat{\boldsymbol{s}})&= \arg \underset{\mathcal{I},\boldsymbol{s}}{\min} \left \| \boldsymbol{y} - \boldsymbol{H} \boldsymbol{x} \right \|^2 \nonumber \\
    &=\arg \underset{\mathcal{I},\boldsymbol{s}}{\min} \left \| \boldsymbol{y} - \boldsymbol{H} \sum_{k=1}^{n_{tx}} s_k \boldsymbol{e}_{I_k} \right \|^2.
\end{align}
The spectral efficiency in terms of bits per channel use (bpcu), is formulated as
\begin{equation}
\label{eq-FA-IM-SE}
\mathrm{SE}= n_{tx} \log_2 Q_m + \left \lfloor \log_2 \binom{N_{tx}}{n_{tx}} \right \rfloor  \quad [\mathrm{bpcu}].
\end{equation}%
The $\left \lfloor \log_2 \binom{N_{tx}}{n_{tx}} \right \rfloor$ part of~\eqref{eq-FA-IM-SE} is  the additional spectral efficiency gained from IM.
%%%%%%%%%%%%%%%%%%%%%%%%%%%%%%%%%%%%%%%%%%%%%%%%%%%%%%%%%%%%%%%%%%%%%%%%%%%%%%%%

\subsection{Multi-user FAMA and its Variants} % Hanjiang
\subsubsection{Transmission Model}
As shown in Fig.~\ref{fig_mu_homo}(b), the received signal of at the $n$-th port of user $u$ is modelled as
\begin{equation}
r_n^{(u)} = h_n^{(u,u)} s^{(u)} + \sum_{\substack{\tilde{u}=1\\\tilde{u}\neq u}}^{U} h_n^{(\tilde{u},u)} s^{(\tilde{u})} + \eta_n^{(u)},
\end{equation} 
where $h_n^{(\tilde{u},u)}$ is the channel from the $\tilde{u}$-th BS antenna to UE $u$ at the $n$-th port, $\eta_n^{(u)}$ is the AWGN with variance of $\sigma_\eta^2$, and $s^{(u)}$ is the transmitted symbol for UE $u$ with $\mathbb{E} [|s^{(u)}|^2] = \sigma_s^2$.

The key point of FAMA is to select the antenna port(s) with the highest SINR at each UE for multiple access, i.e., 
\begin{equation}\label{Eq:PortSelect}
n^* = \arg \max_{n} \gamma_n,~ n = 0, \dots, N^* -1 \text{,}
\end{equation}
where $\gamma_n$ is the SINR metric of the $n$-th port, whose definition varies between \emph{s}-FAMA and \emph{f}-FAMA. 

In \emph{s}-FAMA~\cite{H6_wong2023sFAMA}, $\gamma_n$ is the average SINR calculated as
\begin{equation}
\gamma_n^{{\rm{s}-FAMA}} = \frac{\sigma_s^2 \lvert h_n^{(u,u)}\rvert ^2}{\sigma_s^2 \sum_{\substack{\tilde{u}=1\\\tilde{u}\neq u}}^{U} \lvert h_n^{(\tilde{u},u)}\rvert ^2 + \sigma_\eta^2}.
\end{equation}
The outage probability of \emph{s}-FAMA is thus defined as
\begin{equation}\label{Eq:pout_sFAMA}
  p_{\rm s-FAMA} \!\triangleq\! {\rm Prob} \!\left(\! \max_{n} \frac{\sigma_s^2 \lvert h_n^{(u,u)}\rvert ^2}{\sigma_s^2 \sum_{\substack{\tilde{u}=1\\\tilde{u}\neq u}}^{U} \lvert h_n^{(\tilde{u},u)}\rvert ^2 + \sigma_\eta^2} < \gamma \!\right)\!,
\end{equation}
where $\gamma$ now denotes the SINR threshold.

For \emph{f}-FAMA~\cite{H4_wong2022FAMA,H5_wong2023fast}, on the other hand, $\gamma_n$ is the instantaneous SINR calculated as
\begin{equation}
  \gamma_n^{{\rm {f}-FAMA}} = \frac{\lvert h_n^{(u,u)} s^{(u)}\rvert ^2}{\lvert \tilde{h}_n^{(u)} \rvert ^2}.
\end{equation}
where $\tilde{h}_n^{(u)}\triangleq \sum_{\substack{\tilde{u}=1\\\tilde{u}\neq u}}^{U} h_n^{(\tilde{u},u)} s^{(\tilde{u})} + \eta_n^{(u)}$ represents the instantaneous data-dependent interference and noise. The outage probability of \emph{s}-FAMA can be defined as
\begin{equation}\label{Eq:pout_fFAMA}
  p_{\rm f-FAMA} \!\triangleq\! {\rm Prob} \left(\! \max_{n} \frac{\lvert h_n^{(u,u)} s^{(u)}\rvert ^2}{\lvert \tilde{h}_n^{(u)} \rvert ^2} < \gamma \!\right)\!.
\end{equation}

The multiplexing gain that measures the capacity scaling is an important performance indicator in FAMA. Assume a fixed rate is transmitted to every user, the multiplexing gain in FAMA can be defined as
\begin{equation}
  m_{\rm FAMA} = (1-p_{\rm FAMA})U,
\end{equation}
where $p_{\rm FAMA}$ is given by~\eqref{Eq:pout_sFAMA} or~\eqref{Eq:pout_fFAMA}.

In the simplified channel model~\eqref{model1}, if $|\mu_2| = \dots = |\mu_N| = \mu$, the outage probability of \emph{s}-FAMA is upper bounded by~\cite[Theorem 3]{H6_wong2023sFAMA}
\begin{equation}
  p_{\rm s-FAMA} \!<\! \left[ 1-N\left(\frac{\mu^2}{\gamma+1}\right)^{U-1} - N\left(\frac{1-\mu^2}{\gamma}\right)^{U-1} \right]^{+},
\end{equation}
where the operation $(a)^{+} = \max\{0,a\}$ is to ensure that the bound is never negative. Hence, the multiplexing gain of the \emph{s}-FAMA network, $\color{blue} G_M$, is bounded by~\cite[Theorem 4]{H6_wong2023sFAMA}
\begin{equation}
  U \geq {\color{blue}G_M} \geq \min \left\{ NU\left(\frac{1-\frac{1}{\pi W}}{\gamma}\right)^{U-1},U \right\}.
\end{equation}
In \emph{f}-FAMA, the outage probability is upper bounded by~\cite[Theorem 3]{H4_wong2022FAMA}
\begin{align}
  p_{\rm f-FAMA} \!\lesssim & \frac{1-\mu^2}{4\mu^2}\exp{\frac{1-\mu^2}{4\mu^2}} \nonumber\\ & \times \!\! \sum_{n = 0}^{N-1} \frac{\binom{N-1}{n}(-1)^{n}}{\left(\frac{\sigma_I^2\sigma_s^2\gamma}{\sigma^2}\right)^{n}} e^{\frac{n}{4}}E_k\left(\frac{n}{4}+\frac{1-\mu^2}{4\mu^2}\right),
\end{align}
where $E_k(\cdot)$ represents the generalized exponential integral, $\sigma_I^2 = \sum_{\substack{\tilde{u}=1\\\tilde{u}\neq u}}^{U} \mathbb{E} [|h^{(\tilde{u},u)}_n|^2]\sigma_s^2$ is the variance of interference. Accordingly, the multiplexing gain of the \emph{f}-FAMA network is bounded by~\cite[Corollary 2]{H4_wong2022FAMA}
%\begin{equation}
\begin{align}
  U \geq {\color{blue}G_M} & \gtrsim \min \left\{ \frac{(N-1)(1-\mu^2)U}{\left( \frac{\sigma_I^2 \sigma_s^2 \gamma}{\sigma^2} \right)}, U\right\} \nonumber\\ & \approx \min \left\{ \frac{(N-1)(1-\mu^2)}{\gamma}, U\right\}.
\end{align}
%\end{equation}

However, the channel model~\eqref{model1} is not quite accurate and may lead to artificially optimistic performance predictions as analyzed in Section~\ref{subsec:chan}. Nevertheless, the derivation in the more accurate channel model~\eqref{h_jkn} is much more complicated. The outage probability of the two-user FAMA case in the more accurate channel model~\eqref{h_jkn} was derived in~\cite{H8_Xu2024revisiting}, and~\cite{H7_Espinosa2024Anew} derived the outage probability of \emph{s}-FAMA. 

\begin{figure}
  \centering
  \includegraphics[width = \linewidth]{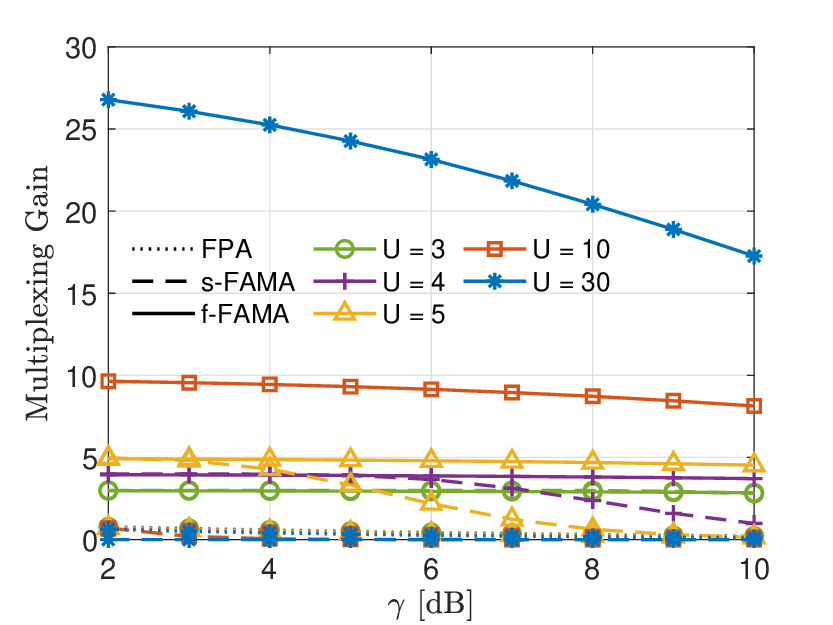}
  \caption{The multiplexing gain for FAMA against the SINR threshold, $\gamma$, for different $U$, when $N = 20\times 20$ and $W = 5\lambda \times 5\lambda$.}\label{Fig:MGvsSINR_FAMA}\vspace{-2mm}
\end{figure}

\subsubsection{Theoretical Performance}
Results in \figref{Fig:MGvsSINR_FAMA} are provided for the multiplexing gain of FAMA compared to FPA systems. It can be seen that the multiplexing gain decreases with the increasing SINR threshold, $\gamma$. The decrease of the multiplexing gain is more prominent with more users (i.e., with a higher $U$). \emph{s}-FAMA provides higher multiplexing gain than FPA when the number of users, $U$, is relatively low. But when $U$ increases, e.g., $U = 10$ or $30$, the multiplexing gain of \emph{s}-FAMA decreases and approaches $0$ because of too much interference. In contrast, \emph{f}-FAMA provides considerable multiplexing gain even in the case of a high number of users, $U$.

\subsubsection{Link-level Simulation Results}
\begin{figure}[t]
  \centering
  \includegraphics[width=\linewidth]{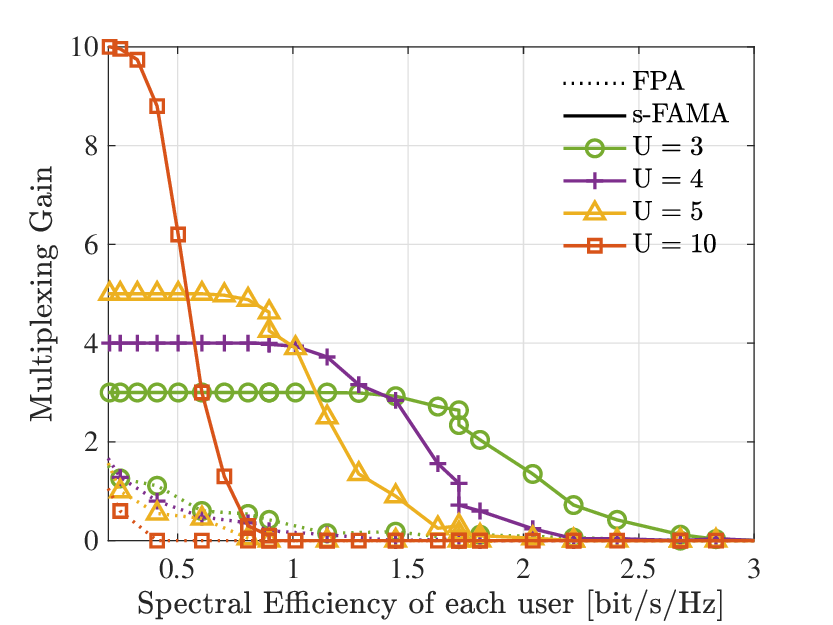}
  \caption{The multiplexing gain for FAMA against the spectral efficiency of each users, when $N = 20\times 20$ and $W = 5\lambda \times 5\lambda$.}\label{Fig:MGvsSE_FAMA}\vspace{-2mm}
\end{figure}
In~\cite{H11_hong2025Downlink}, the link-level performance of \emph{s}-FAMA with consideration of channel coding and OFDM was reported. Link-level simulation results in \figref{Fig:MGvsSE_FAMA} are presented for the multiplexing gain of \emph{s}-FAMA compared with FPA systems. The multiplexing gain in the link-level simulations is re-calculated as ${\color{blue}G_{\rm FAMA}} = (1-{\rm BLER})U$, where ${\rm BLER}$ is the block error rate in the simulations. As we can see, \emph{s}-FAMA provides great multiplexing gain in the low spectral efficiency region. The multiplexing gain decreases with increasing spectral efficiency. This is consistent with the theoretical performance in \figref{Fig:MGvsSINR_FAMA}, since higher spectral efficiency requires higher SINR threshold $\gamma$. Specifically, when $U = 10$, \emph{s}-FAMA can achieve multiplexing gain of $10$ at a very low spectral efficiency, which is not achieved in Fig.~\ref{Fig:MGvsSINR_FAMA}. The reason is that the operating SINR is lower than $\gamma = 2~{\rm dB}$ and not presented in Fig.~\ref{Fig:MGvsSINR_FAMA} for this low spectral efficiency.

\subsubsection{Variants of FAMA}
While considering the case of multiple activated ports,~\cite{H12_Wong2024cuma} proposed the CUMA scheme that processes the in-phase and quadrature parts of the receieved symbols individually. In CUMA, the activated ports can be chosen to add constructively to strengthen the desired signal. Thus, the classification of the ports can be performed as~\cite{H12_Wong2024cuma}
\begin{equation}\label{Eq:cuma_cri}
  \left| \sum_{n \in \mathcal{N}^{+}} {\rm real}(h^{(u,u)}_n)\right| {\substack{\mathcal{N}^{+} \\ > \\ < \\ \mathcal{N}^{-}}} \left| \sum_{n \in \mathcal{N}^{-}} {\rm real}(h^{(u,u)}_n)\right|,
\end{equation}
in which $\mathcal{N}^{+}$ denotes the set of port indices for all positive in-phase channels and $\mathcal{N}^{-}$ for the negative in-phase channels. With a predefined minimum required level of the in-phase channel to be on, $\rho$, the $n$-th port can be selected only if 
\begin{equation}\label{Eq:cuma_rho}
  {\rm real}(h^{(u,u)}_n) \geq \rho \max_{m \in \mathcal{K}^{s}} {\rm real}(h^{(u,u)}_m), ~ s \in \{+,-\}.
\end{equation}
The final set, $\mathcal{N}$, can be obtained by choosing from $\mathcal{N}^+$ and $\mathcal{N}^-$ using the criterion~\eqref{Eq:cuma_cri} but only those ports satisfying~\eqref{Eq:cuma_rho} is summed. Then, the in-phase and quadrant parts of the received symbol can be aggregated as~\cite{H12_Wong2024cuma}
\begin{equation}
  \left\{\begin{aligned}
    & r_u^{I} = \sum_{k \in \mathcal{N}} {\rm real}(r^{(u)}_n), \\
    & r_u^{Q} = \sum_{k \in \mathcal{N}} {\rm imag}(r^{(u)}_n).
  \end{aligned} \right.
\end{equation}
Finally, the received symbol is detected as~\cite{H12_Wong2024cuma}
\begin{equation}
  \tilde{s}_u \!\!=\!\! \left[\!\!\!
    \begin{array}{lr}
      \sum_{n \in \mathcal{N}} \!{\rm real}(h^{(u,u)}_n) \!\!&\!\! -\sum_{n \in \mathcal{N}} \!{\rm imag}(h^{(u,u)}_n)\\
      \sum_{n \in \mathcal{N}} \!{\rm imag}(h^{(u,u)}_n) \!\!&\!\! \sum_{n \in \mathcal{N}} \!{\rm real}(h^{(u,u)}_n)
    \end{array} \!\!\!\right] \!\!\times\!\!
    \left[\!\!\! \begin{array}{l}
      r_u^I\\ r_u^Q
    \end{array}\!\!\!\right].
\end{equation}
The quality of detection can be further improved by repeating the above-mentioned process with focusing on the quadrature component. With the sophisticated selection of the ports, CUMA can support hundreds of users per channel use.
%%%%%%%%%%%%%%%%%%%%%%%%%%%%%%%%%%%%%%%%%%%%%%%%%%%%%%%%%%%%%%%%%%%%%%%%%%%%%%%%%%%%%%%%%%%%%%%%%%%%%%%%%%%%%%%%%%%%%%%%%%%%%%%%%%%%%%%%%%%%%%%%%%%%%%%%%%%%%%%%

\begin{figure*}
\centering
\subfigure[]{\includegraphics[width = .4\linewidth]{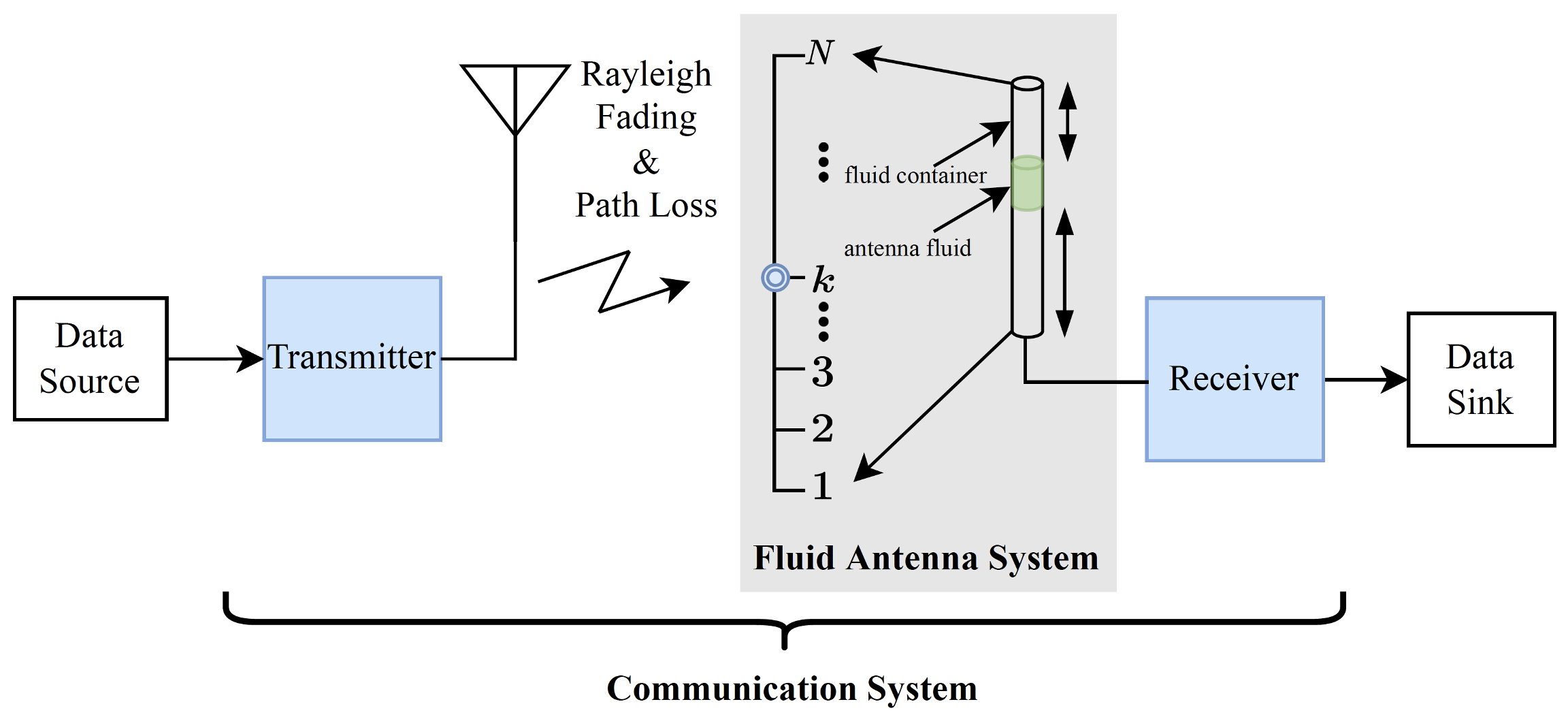}\label{Fig9a_qFAS}}%\vspace{-2mm}
\subfigure[]{\includegraphics[width = 0.6\linewidth]{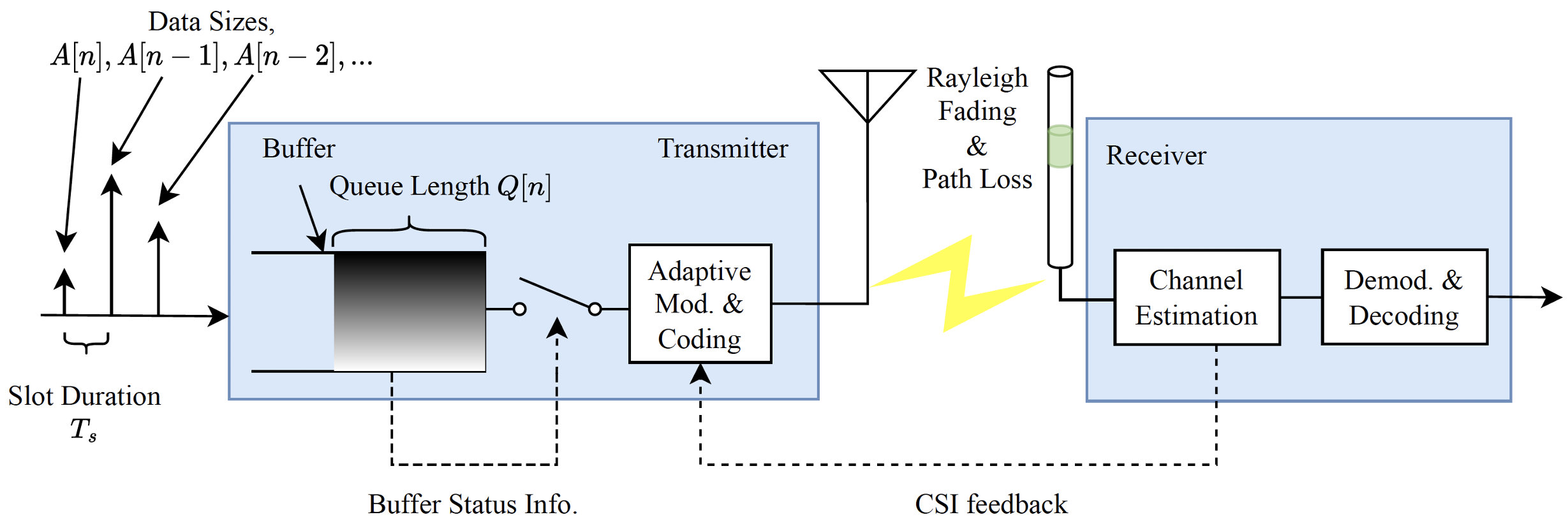}\label{Fig9b_qFAS}}%\vspace{-2mm}
\caption{The communication system model for a SISO-FAS: (a) a point-to-point channel model and (b) a discrete-time queueing model for FAS. Note that even a fluidic FAS is shown but in practice, it could be realized using other technologies such as pixels and metamaterials.}\vspace{-2mm}
\end{figure*}

\section{Networking Technology}\label{sec:network}
While the foundational techniques of the physical layer establish the basis for FAS, networking techniques are equally essential for ensuring the efficient, reliable, and secure operation of practical implementations. Given that research on FAS is relatively recent, investigations into their upper layers remain in the preliminary stages. Nevertheless, the impact of networking techniques on the performance capabilities of FAS has increasingly garnered significant interest from both academic and industry sectors. 

One of the primary objectives for networking technologies is the effective exploitation of radio resources. In this context, EE is considered as a critical performance metric. Moreover, it is essential to guarantee superior QoS within the constraints of limited radio resources. This section focuses on the QoS enhancement in the FAS network, exploring power allocation strategies under QoS constraints, and studying content placement methodologies in content-centric FAS HetNets.

\subsection{QoS Enhancement} % Yu
% {\color{blue}QoS enhancement technology refers to a suite of networking techniques and mechanisms designed to manage, prioritize, and optimize network traffic to ensure that critical applications receive the necessary bandwidth, low latency, low jitter, and minimal packet loss they require, even when the network is congested. The key techniques of QoS enhancement include classification and marking, congestion management, traffic shaping and policing, and congestion avoidance. Classification and marking identify different types of teletraffic and mark packets with a priority level as they enter the network.  Congestion management, also known as queuing, is used to determine which packets are sent first when an interface is congested. It can prioritize high-priority traffic by giving it preferential access to the outgoing link, ensuring minimal delay and jitter. Traffic shaping and policing control the rate of traffic sent into or out of the network. Specifically, traffic shaping buffers excess teletraffic and smooths out burst to conform to a defined rate, while traffic policing drops packets that exceed the defined rate. The congestion avoidance proactively monitors network queues and signals senders to slow doen before queues fill up and packets are dropped.}

{\color{blue}QoS enhancement techniques are mechanisms that improve the quality of data transmission, particularly in terms of packet delay and packet loss ratio, which are two key performance indicators.}
{\color{blue}FAS dramatically enhances QoS through spatial diversity compression, enabling hundreds of antenna ports within a compact space~\cite{tle2022enhancing}. Thus, it is important to reconsider the statistical behavior of FAS for QoS enhancement in the network. Packet delay is one QoS metric considered in many wireless communication networks. For this reason, we focus on this delay to preliminary discuss the QoS enhancement in FAS network below.}

\subsubsection{QoS Modelling in Mobile Networks}
Fig.~\ref{Fig9a_qFAS} depicts a simplified FAS-based point-to-point communication system model from an upper-layer data source to a data sink. With FAS installed at the receiver side, the received signal can be denoted as in~\eqref{eq:siso-fas}, with the channel model specified in Section~\ref{subsec:chan}. Moreover, the communication system inside the bracket of Fig.~\ref{Fig9a_qFAS} is termed an FAS-based communication system; the details of such a system at slot $n$ is shown in Fig.~\ref{Fig9b_qFAS}.

The communication system in Fig.~\ref{Fig9a_qFAS} can be described as a discrete-time queueing model in Fig.~\ref{Fig9b_qFAS}. Let us denote the system bandwidth by $B$, the transmit power by $P_{\rm t}$, and the noise spectral density by $N_0$. This system is discrete-time; the slot duration is $T_s$. Provided that a path loss is $L_p$ and the noise variance is given by $\sigma_\eta^2 = N_0 B$~~\cite{Ozcan2018}, the average SNR, denoted by $\Gamma$, is defined as
\begin{equation} \label{deqn_ex3}
  \Gamma = \sigma^2 \frac{\mathbb{E}\left[|s|^2\right]}{\sigma_\eta^2}
  =\sigma^2 \frac{ P_{\rm t}}{N_0 B}
  \xrightarrow[\sigma^2]{L^{-1}_p} \frac{1}{L_p} \frac{ P_{\rm t}}{N_0 B} = \frac{ P_{\rm t}}{L_p \sigma_\eta}.
\end{equation}
  
Upper-layer data from the data source are first pushed into a first-in-first-out (FIFO) buffer at the transmitter, and then transmitted to the receiver over a block-fading channel (i.e., the channel gains $\{g_k\}_{\forall k}$ remain constant during a time slot). Specifically, we define the following variables:
\begin{enumerate}
\item[${A[n]}$] for the amount of data in bits from the data source at time slot $n$. The RVs $A[1], A[2], \dots $ are assumed to be i.i.d.~exponential random variables with its probability density function (PDF) being
\begin{equation}\label{eq:fA}
f_A(a)=\begin{cases}
\lambda \exp \left( -{\lambda a}\right) & \mbox{for }a \ge 0,\\
0 & \mbox{for }a < 0.
\end{cases}
\end{equation}
The average data rate is $R  = (T_s \lambda)^{-1}$ (bps);
\item[${S[n]}$] for the amount of data in bits that the transmitter is capable of transmitting at slot $n$. The random variables $S[1], S[2],\dots $ are i.i.d.~from some fixed distribution;
\item[$Q{[n]}$] for the backlog size in bits at slot $n$.
\end{enumerate}
  
With the instantaneous channel gain $h_{\rm FAS}$ perfectly known at the transmitter side, the FAS-based system capacity $C$ at slot $n$ could be approximated as
\begin{equation}\label{Eq:c_n}
C \approx B \log_2\left(1+ \frac{h_{\rm FAS} P_{\rm t}}{N_0 B} \right) =
B \log_2\left(1+ \frac{h_{\rm FAS} P_{\rm t}}{\sigma_\eta^2} \right).
\end{equation}
The capacity $C$ and the service $S$ have a simple relation 
\begin{equation}
S = C T_s.
\end{equation}
  
\subsubsection{Delay Distribution as QoS Metrics}
An analytical expression of delay distribution can be obtained by the use of the effective bandwidth and the effective capacity models. Because $A$ is an exponential random variable with its PDF~\eqref{eq:fA}, the effective bandwidth of the arrival process is given by~\cite{Chen2015a}
\begin{align}
      \alpha^{(b)}(u) = \frac{1}{T_s u} \log\left(\frac{\lambda}{\lambda - u}\right),
\end{align}
where $u$ is the QoS exponent.
  
The PDF of the channel power gain $g_{\rm FAS}$ in an $N$-port FAS based on the simplified model~\eqref{model1} is given by
\begin{equation}\label{eq:hk}
f_{h_{\rm FAS}}(x) =
  \begin{cases}
  \frac{1}{\sigma^2}\exp \left(-\frac{x}{\sigma^2} \right), & N=1 \\[8pt]
  \frac{N}{\sigma^2} \int_{0}^{\infty} F_{X}(x|r)^{N-1} f_{X}(x|r) e^{-\frac{r}{\sigma^2}} \, dr, & N>1
  \end{cases}
\end{equation}
where
\begin{align}\label{eq:Fx}
    F_{X}(x|r) = 1 - Q_1\left(\sqrt{\frac{\mu^2 r}{\sigma^2(1-\mu^2})}, \sqrt{x} \right),
\end{align}
and
\begin{align} \label{eq:fx}
f_{X}(x|r)
  &= \frac{1}{2}e^{ - \frac{{x + \frac{{{\mu ^2} r}}{\sigma^2({1 - \mu ^2})}}}{2}}{J_0}\left(\sqrt {\frac{{{\mu ^2} h_0 x}}{\sigma^2({1 - {\mu ^2})}}} \right).
\end{align}
Based on this PDF, the effective capacity is found as~~\cite{Chen2015a}
\begin{align}
  \begin{aligned}
  &\alpha^{(c)}(u;P_{\rm t}) \\
  &= \frac{-1}{T_s u} \log \int_0^{\infty} e^{-u T_s B \log_2(1 + x P_t / \sigma_\eta)} f_{h_{\rm FAS}}(x) d x \\
  &= \frac{-1}{T_s u} \log \int_0^{\infty} \left(1+ \frac{x P_t}{\sigma_\eta}\right)^{-\frac{uT_sB}{\log(2)}} f_{h_{\rm FAS}}(x) d x.
  \end{aligned}
\end{align}

If the assumptions of the Gartner-Ellis theorem hold and if there is a unique QoS exponent $u^*>0$ that satisfies
\begin{align} \label{eq:ebec}
  \alpha^{(b)}(u^*)=\alpha^{(c)}(u^*;P_{\rm t}),
\end{align} 
the complementary cumulative distribution function (CCDF) of delay can be accurately approximated by~~\cite{Chen2015a}
\begin{align}\label{eq:dop}
\begin{aligned}
S_N(P_{\rm t};D_{\rm max}) &= {\rm Prob}(D(\infty) > D_{\rm max}) \\
&\approx {\left( {1 - \lambda^{-1}{u^*}}\right)^{D_{\rm max} + 1}},~\mbox{for }D_{\rm max} \in \mathbb{N}_0,
\end{aligned}
\end{align}
where $\mathbb{N}_0$ is a set of natural numbers including zero. Also, the expectation and variance of delay can be approximated as
\begin{align}\label{eq:mean}
  \mathbb{E}[D(\infty)]=\frac{1 - \lambda^{-1}{u^*}}{ \lambda^{-1}{u^*}}
\end{align}
and
\begin{align} \label{eq:var}
  {\rm Var}[D(\infty)]=\frac{1 - \lambda^{-1}{u^*}}{ (\lambda^{-1}{u^*})^2}.
\end{align}
  
To illustrate the results, we simulate an $N$-port FAS with the path loss $L_p$ based on the 3GPP model for a carrier frequency between $1400$ MHz and $2600$ MHz~~\cite{3GPP2022}, defined as
\begin{align}
{L_p} = 128.1 + 37.6{\log _{10}}\left( d \right) + 21 \log_{10}\left(\frac{f_c}{2}\right),
\end{align}
where $d$ denotes the distance between the transmitter and the receiver in $\rm km$ and $f_c$ is the carrier frequency in ${\rm GHz}$. In the simulations, we fix the distance $d$ to $0.2~{\rm km}$, $f_c$ to $2~{\rm GHz}$, the bandwidth to $10~{\rm MHz}$ and the constant circuit power $P_c$ to $0.1$ Watts, the noise spectral density $N_0$ to $-174~{\rm dbm/Hz}$ and the slot duration to $1~{\rm ms}$.
  
\begin{figure}[!t]
\center{\includegraphics[width=\linewidth]{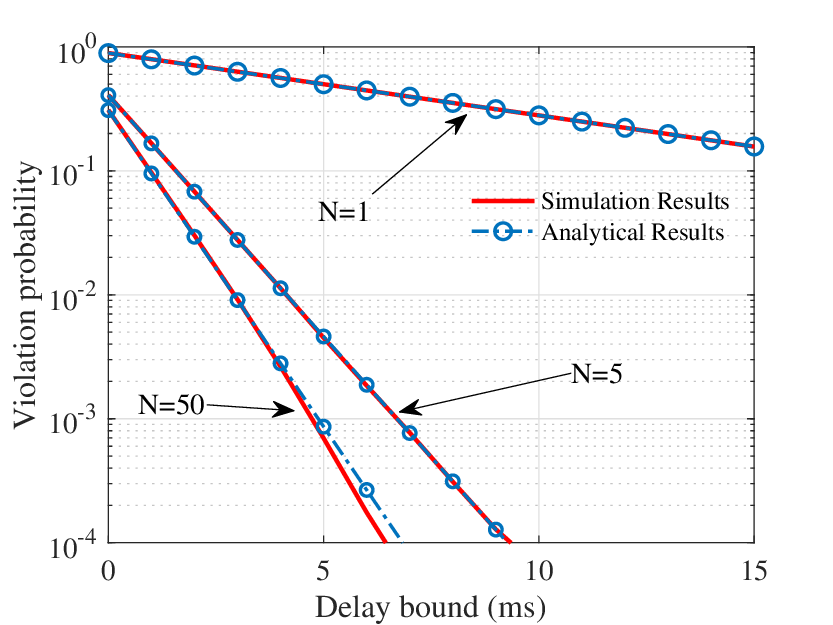}}
\caption{Simulation and approximation results of ${\rm Prob}(D > D_{\rm max})$.} \label{fig:ccdf_vs_np}\vspace{-2mm}
\end{figure}

Consider a 1D-FAS with $W=1\lambda$, which is about $0.15~{\rm m}$. Given that the transmit power is $20~{\rm dBm}$ and the average data rate $R$ is $60~{\rm Mbps}$, we show the simulation and analytical results of the delay-outage probability (DOP) with respect to different values of the maximum delay bound $D_{\rm max}$ in Fig.~\ref{fig:ccdf_vs_np} when $N=1, 5$ and $50$. The $x$-axes are delay bounds (the unit is millisecond) and the $y$-axes are violation probabilities in log scale. Under any number of ports, the simulation result is obtained based on $10^6$ samples of backlog size ($1000$ seconds $\approx 0.277$ hours), the QoS exponent $u^*$ is numerically found based on an algorithm in~\cite{Chen2015a} and the analytical result is calculated by~\eqref{eq:dop}. As illustrated in Fig.~\ref{fig:ccdf_vs_np}, the simulation and analytical results almost overlap with each other in all three scenarios. This implies that (\ref{eq:dop}) can accurately approximate the CCDF of delay in FASs with exponential arrival processes. On the other hand, the delay performance can be substantially improved by increasing the number of ports.
  
\subsection{\color{blue}Power Allocation}
{\color{blue}The power allocation technology distributes transmit power across multiple users, channels, antennas, or subcarriers in a communication network. Its goal is to maximize overall system performance, such as data rate, coverage, or capacity, while adhering to total power constraints and often considering fairness between users. Power allocation in FAS involves determining optimal power distribution across users or spatial streams while leveraging FAS's ability to dynamically select optimal FAS ports to maximize system objectives. The unique characteristics of FAS, such as high spatial-domain flexibility and CSI dependency, should be considered when allocating power. In this section, we summarize the power allocation technology in FAS from the perspectives of optimization objectives, solution methodologies, and its integration with emerging applications. We also include a special case of power allocation under QoS constraints to help readers better understand this technology.}
\subsubsection{\color{blue}Optimization Objectives and Constraints}
{\color{blue} Power allocation in FAS targets diverse objectives subject to application-specific constraints, mainly including EE maximization and sum-rate maximization. The EE maximization problem is usually considered under delay-outage probability constraints~\cite{li2024EE,xu2023EE}. FAS enhances EE by exploiting spatial diversity to reduce transmit power while maintaining QoS. The statistical delay-bound, total power budget, and FAS port activation limits are the common constraints in EE maximization problem. In multi-user FAS networks, sum-rate maximization is a promising technique to achieve higher aggregate data rates. FAS enhances sum-rate by suppressing multi-user interference through optimal port selection, enabling power allocation to focus on signal enhancement rather than interference mitigation~\cite{tang2024power,abdou2024sumrate}. Minimum rate guarantees are the basic constraint in this problem.}
\subsubsection{\color{blue}Solution Methodologies}
{\color{blue}The solution methodologies for FAS power allocation can be classified into the convex optimization and decomposition, learning-based approaches, and hybrid techniques. With convex optimization and decomposition solution, the non-convex joint optimization FAS power allocation problem can be decomposed into tractable problems, such as large-scale or small-scale fading optimization. Deep reinforcement learning, on the other hand, is suitable to address high-dimensional state spaces of the large number of FAS ports and CSI~\cite{ho2025deep}. However, it may face challenges related to sample inefficiency and difficulties in designing a reward function for multi-objective optimization.}
\subsubsection{\color{blue}Integration into Emerging Applications}
{\color{blue}More recently, FAS power allocation techniques in emerging application scenarios have been proposed. For instance,~\cite{abdou2024sumrate} studied the architecture where unmanned aerial vehicle (UAV) relays facilitate line-of-sight links to FAS-equipped ground users. FAS mitigates small-scale fading, while NOMA enables multiplexing. The power allocation in this scenario jointly optimizes the UAV altitude, FAS port selection, and NOMA power coefficients. In the energy harvesting scenario, FAS ports co-located with RF energy harvesters enable simultaneous exploitation of spatial diversity and energy focusing. The simultaneous wireless information and power transfer optimization problem balances the tradeoffs by allocating power across the information-bearing signals and energy-carrying signals \cite{tong2025design}.}
\subsubsection{\color{blue}Case Study: Power Allocation under QoS Constraints} % Yu
Consider that the total transmission power can be expressed as $P_{\rm tot}=P_c+P_{\rm t}$, where $P_c$ is the constant circuit power and $P_{\rm t}$ is the transmit power consumed by power amplifier. Based on Zorzi and Rao's definition in~\cite{Zorzi1997}, an $N$-port FAS's EE can be found as
\begin{align} \label{eq:ee}
\eta_N(P_{\rm t}) =\frac{R}{P_c+P_{\rm t}}.
\end{align}

As shown in Fig.~\ref{Fig9b_qFAS}, upper-layer data experience delays in a queueing buffer. We first define a maximum delay bound $D_{\rm max}$ and a tolerance $\epsilon$, and specify a target DOP constraint $\{{D_{\max }},\epsilon\}$ in order to support QoS for a hypothetical application. The system has to guarantee the following inequality:
\begin{align}\label{eq:10}
  {\rm Prob}\left( {D(\infty) > {D_{\max }}} \right) \le \epsilon,
\end{align}
where $D(\infty)$ is the steady-state delay. Given that the transmitter decides the allocation of the transmit power $P_{\rm t}$ on the basis of the constraint $\{{D_{\max }},\epsilon\}$ and the rest parameters are fixed, the EE can be optimized. That is to solve:
\begin{align}
\max_{P_{\rm t} \ge 0} &~~ \eta_N(P_{\rm t})\\
\text{s.t.} &~~ {\rm Prob}(D(\infty) > D_{\rm max}) \le \epsilon. \nonumber
\end{align}

\begin{figure}[!t]
\center{\includegraphics[width=\linewidth]{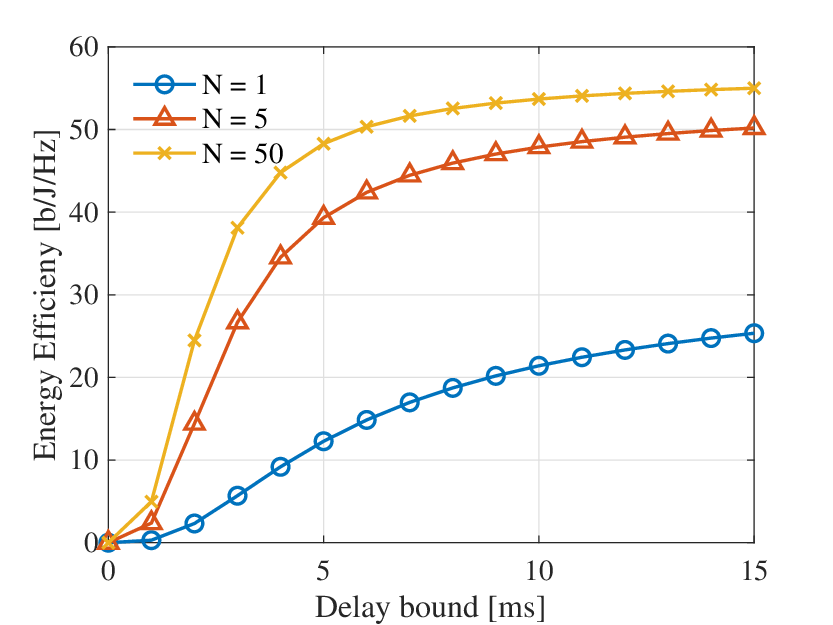}}
\caption{Optimal EE under different DOP and number of ports of FAS.} \label{fig:ee_vs_dmax}\vspace{-2mm}
\end{figure}
  
Given the DOP constraint $\{D_{\rm max}, \epsilon\}$ and that the average packet length $1/\lambda$ at a slot is known, the target QoS exponent $u^{\dag}$ can be expressed as
\begin{align} \label{eq:target_u}
    u^{\dag} = \lambda \left(1 - \epsilon^{\frac{1}{D_{\rm max} + 1}}\right),
\end{align}
which is an inverse of the function~\eqref{eq:dop} with respect to $D_{\rm max}$ and $\epsilon$. Moreover, the optimal power $P^*_{\rm t}$ is unique and satisfies the following equation:
\begin{align}\label{eq:ebecp}
  \lambda\left( \int_0^{\infty} \left(1+ \frac{x P_{\rm t}}{\sigma_\eta}\right)^{-\frac{u^{\dag} T_sB}{\log(2)}} f_{g^*}(x; P_{\rm t}) d x - 1\right) = u^{\dag},
\end{align}
and $P^*_{\rm t}$ can be solved by any root-finding numerical methods, e.g., binary search method.
  
Consider again the same FAS with $W = 1\lambda$. In Fig.~\ref{fig:ee_vs_dmax}, we demonstrate numerical results on the maximum achievable EE obtained from~\eqref{eq:ebecp} for different DOP constraints, i.e., $D_{\rm max}=\{0, 1, 2, \dots, 15\}$ ms where $\epsilon=0.02$. We see that EE increases when extending $D_{\rm max}$ and relaxing the DOP. When $D_{\rm max}$ goes to infinity, the optimal transmit power converges a value that satisfies $R=\mathbb{E}[C]$, which is the condition for a stable system. Again, the EE in FAS-based communication systems is significantly higher compared with single-antenna FPA scenario. In the case of $D_{\rm max}=1\rm{~or~}10~ms$ (required for 5G/6G networks~~\cite{series2015imt}, respectively), the EE in a 50-port FAS is about $2.02$ to $16.11$ times higher than a FPA system.
  
\begin{figure}[!t]
\center{\includegraphics[width=\linewidth]{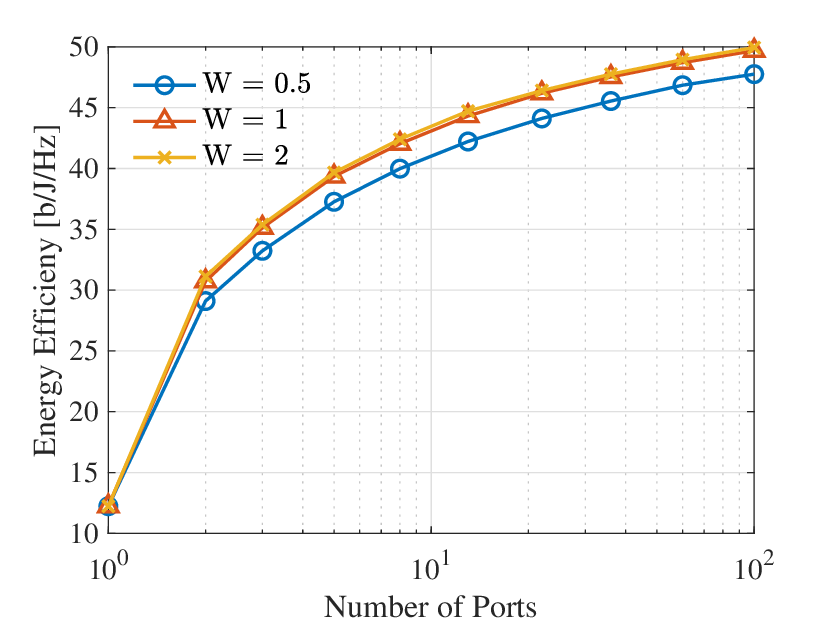}}
\caption{Optimal EE using different number of ports of FAS and $W$ under a DOP constraint $\{D_{\rm max} = 5~\rm{ms}, \epsilon = 0.02\}$.}\vspace{-2mm}\label{yu_fig9}
\end{figure}

Fig.~\ref{yu_fig9} shows numerical results on the optimal EE under a target DOP constraint $\{D_{\rm max} = 5~\rm{ms}, ~\epsilon = 0.02\}$ using different number of ports, $N$, and different normalized size of FAS, $W$. The EE increases as $N$ increases, but this growth becomes less pronounced when $N$ is large. Additionally, it can be observed that increasing the size of FAS does enhance the EE of the systems, but the improvements are marginal.

%%%%%%%%%%%%%%%%%%%%%%%%%%%%%%%%%%%%%%%%%%%%%%%%%%%%%%%%%%%%%%%%%%%%%%%%%%%%%%%%
\subsection{\color{blue}Network Technologies in FAS HetNets}
%{\color{blue} FAS exploits spatial degrees of freedom to adapt to channel conditions, offering unprecedented diversity gains. In HetNets, FAS addresses critical challenges like interference management, mobility robustness, and spectral efficiency. With future 6G networks demanding massive connectivity and ultra-reliability, FAS emerges as a transformative technology for enhancing HetNet performance. The flexibility of FAS will introduce additional spatial diversity in HetNets, which should be reconsidered for network technology development. For instance, FAS reduces the handover frequency since it can maintain connectivity in ultra-dense networks. The handover thresholds need to be reconsidered with the assistance of FAS. However, the network technology study of FAS-assisted HetNets remains its infancy. Here we focus on a special case of content placement in content-centric FAS HetNets.}

% Farshad
%{\color{blue}Content placement technology strategically stores popular or critical content closer to end-users to optimize network performance, enhance user experience, and reduce operational costs. It can reduce latency, save bandwidth, improve reliability, and can be effectively scaled. FAS equipped at UE can improve the content delivery performance. However, the characteristics of FAS should be considered in the content placement in FAS CCN. Thus, the content placement technology in the FAS-equipped CCN is reviewed in this subsection.}
\textcolor{blue}{FAS introduce a novel spatial degree of freedom by allowing antenna elements to dynamically reposition over a predefined surface. This flexibility enhances spatial diversity, improves signal robustness, and provides fine-grained control over the propagation environment. In HetNets, where dense deployments, user mobility, and interference are major challenges, FAS offers promising solutions through dynamic beam steering, adaptive user association, and spatial interference mitigation. Unlike traditional antennas with fixed radiation patterns, FAS can continuously adapt to instantaneous channel conditions, leading to reduced handover frequency and improved mobility robustness. This is particularly beneficial in ultra-dense small cell deployments where conventional handover mechanisms become inefficient. As a result, handover thresholds, association criteria, and beam management policies must be revisited to fully leverage the benefits of FAS.}

\textcolor{blue}{With the advent of 6G networks demanding ultra-reliability, low latency, and massive connectivity, FAS emerges as a transformative technology to enhance HetNet performance. However, its dynamic behavior introduces new complexities: user association becomes less predictable, CSI acquisition and tracking become more frequent, and spatial agility adds a new dimension to scheduling and resource allocation algorithms. A particularly relevant application area is CCN, where popular or delay-sensitive content is proactively cached closer to users to reduce latency and backhaul load. In FAS-enabled CCNs, the spatial agility of FAS-equipped UE and base stations can be exploited to maintain more stable and higher-quality links, thereby improving content delivery performance. However, the dynamic and location-adaptive nature of FAS complicates traditional content placement strategies, which often assume static or slowly varying user association patterns. This calls for new caching and content delivery algorithms that are aware of FAS-specific characteristics such as antenna scanning range, beam alignment latency, and spatial association variability.}

\textcolor{red}{In the remainder of this section, we detail the system model and enabling technologies for FAS-assisted HetNets, with a particular focus on content delivery and placement mechanisms in content-centric architectures.}

\subsubsection{Transmission Model}
As illustrated in Fig.~\ref{Fig_CCN}, considering a multi-user CCN-enabled HetNet setup comprising  SBSs tiers and 2D-FAS-equipped UEs. It is assumed that each SBS has limited storage and caches the most popular content. The UEs connect to their nearest SBS(s) that has the desired content, with a finite set of contents $\mathcal{F} \coloneqq \left\{f_1,\dots,f_l,\dots, f_L\right\}$, where $f_l$ is the $l$-th most popular content, while the SBSs connect to the backhaul via a macro-cell BS (MBS). The popularity of each content follows a Zipf distribution, with the request probability for the $l$-th content given by~\cite{F4_zhu2018content}
\begin{align}
p_l = \frac{l^{-\zeta}}{\sum_{k=1}^{L}k^{-\zeta}}, 
\end{align}
in which $\zeta$ is the Zipf exponent and $\sum_{l=1}^{L}p_l=1$. Each SBS can cache up to $K$ contents, with $K\ll L$. Additionally, a probabilistic caching strategy is assumed, where each content is independently cached at each SBS with probability $q_l$. Thus, the set of caching probabilities $\textit{q}=\left\{q_1,\dots,q_L\right\}$ must satisfy the constraint $\sum_{l=1}^{L} q_l\leq K$. It is also assumed that content is arranged in descending order of request probability, with more popular content being more likely to be cached.

A typical FAS-equipped UE with $N = N_1 \times N_2$ ports and size of $W = W_1 \lambda \times W_2 \lambda$ is located at the origin of a Cartesian plane, while the locations of SBSs are assumed to follow an independent homogeneous Poisson point process (HPPP) $\Psi^\mathrm{S}$ with intensity $\mu_\mathrm{S}$. Specifically,  the point process $\Psi^\mathrm{S}_l$ represents the SBSs that cache content $l$, with intensity $q_l\mu_\mathrm{S}$ in each tier. The intensity of UEs is assumed to be much higher than that of SBSs, and each FAS-equipped UE is allowed to communicate with one SBS per time slot. Therefore, the received SNR at a FAS-equipped UE requesting content $l$ from its respective SBS that includes this content is defined as
\begin{align}\label{eq:snr}
\gamma_l=\frac{P_t\left| h^{(l)}_{n^*}\right|^2 D\left(\vert X_l\vert\right)}{\sigma_\eta^2},
\end{align}
where $P_t$ denotes the transmit power, and $D\left(\vert X_l\vert\right)=\beta\vert X_l\vert^{-\alpha}$ represents the path-loss with the distance $\vert X_l\vert$, in which $\beta$ and $\alpha$ are the frequency-dependent constant and the path-loss exponent, respectively. Under the assumption of advanced interference management algorithms, communications are noise-limited, and inter-cell interference is relatively low, so $\sigma_\eta^2$ in~\eqref{eq:snr} represents the combined power of noise and weak interference at the FAS-equipped UE. Additionally, $n^*$ is the index of the best port, given by
\begin{align}\label{Eq:K_star}
	n^* = \arg\max_n\left\{\left| h^{(l)}_{n}\right|^2\right\},
\end{align}
where $h^{(l)}_{n}$ is the channel coefficient between the $n$-th port of a typical FAS-equipped UE and its serving SBS, and the channel gain $g^{(l)}_{n}=\left| h^{(l)}_{n}\right|^2$ follows exponential distribution, i.e., $g^{(l)}_{n}\sim\mathrm{Exp}\left(1\right)$. According to~\eqref{Eq:K_star}, we have
\begin{align}\label{eq-gain}
g^{(l)}_\mathrm{FAS} = g^{(l)}_{n^*} = \max\left\{g_1^{(l)},\dots,g^{(l)}_N\right\}.
\end{align}

By assuming a 3D environment under rich scattering, the covariance $J_{k,l}$ between two arbitrary ports $k$ and $l$ is characterized as~\eqref{J_tx_ele}. Note that in the FAS-equipped UEs only the optimal port that maximizes the received SNR is activated. Hence, as shown in~\cite{F5_rostami2024gaussian}, the covariance of Jakes' model $J_{k,l}$ is effectively  approximated with the dependence parameter $\vartheta$ of elliptical copulas, e.g., the Gaussian copula, the cumulative density function (CDF) of $g^{(l)}_\mathrm{FAS}$ is given by~~\cite{F5_rostami2024gaussian}
\begin{align}
F_{g^{(l)}_\mathrm{FAS}}\!\left(r\right) \!\! = \!\! \Phi_\mathbf{R} \!\! \left(\! \phi^{-1}\!\left(F_{g_1^{(l)}}\left(r\right)\right)\!,\dots\!,\phi^{-1}\!\left(F_{g_N^{(l)}}\left(r\right)\right);\vartheta \!\right)\!,
\end{align}
where $\Phi_\mathbf{R}$ is the joint CDF of a multivariate normal distribution with mean zero and correlation matrix $\mathbf{R}$, and $\phi^{-1}\left(x\right)=\sqrt 2\mathrm{erf}^{-1}\left(2x-1\right)$ is the quantile function of the standard normal distribution, in which $\mathrm{erf}^{-1}\left(\cdot\right)$ refers to the inverse error function and $F_{g_k^{(l)}}\left(r\right)=1-\mathrm{e}^{-r}$.

\subsubsection{Performance Analysis} 
The performance of content delivery systems is crucial for optimizing network efficiency. Two key performance metrics that are essential for analyzing the performance of CCN are the successful content delivery probability (SCDP) and the content delivery delay (CDD). The SCDP quantifies the likelihood that a content requested by a typical UE is both cached within the network and successfully transmitted, providing an indication of the system's efficiency in meeting user demands. On the other hand, CDD measures the delay associated with delivering content to the UE, taking into account retransmission protocols such as automatic repeat request (ARQ) and the impact of channel fading. 

\begin{figure}[!t]
\centering
\includegraphics[width=\linewidth]{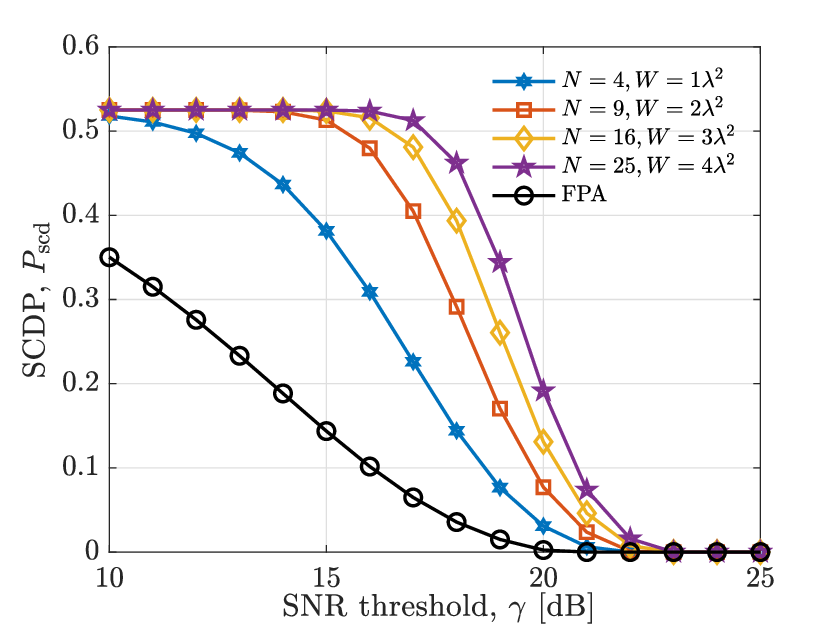}
\caption{SCDP versus threshold $\gamma$ when $q_l=1$.}\vspace{-2mm}\label{fig-scd-eta}
\end{figure}

\begin{figure}[!t]
\centering
\includegraphics[width=\linewidth]{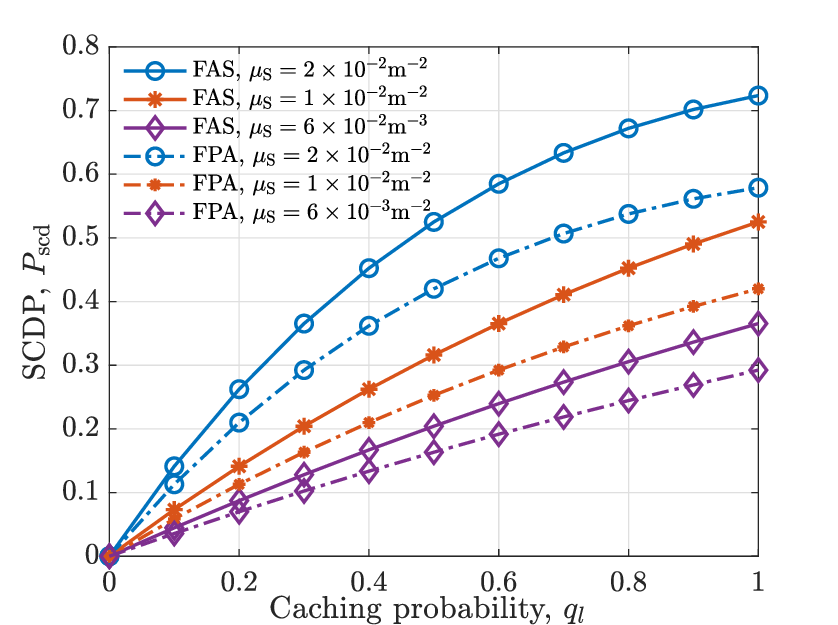}
\caption{SCDP versus caching probability $q_l$ when $N=3\times 3$, $W=1\lambda \times 1\lambda$.}\label{fig-scd-q}\vspace{-2mm}
\end{figure}

For downlink, the SCDP is mathematically defined as $P_\mathrm{scd}=\sum_{l=1}^{L}p_l P_{\mathrm{s},l}$, where the probability of successful transmission is calculated as $P_{\mathrm{s},l}={\rm Prob}\left(\gamma_l\ge\gamma\right)$, and $\gamma$ represents the SNR threshold.  Given that $P_{\mathrm{s},l}$ is conditioned on the random variable $\left|X_l\right|$, which represents the distance between the typical FAS-equipped UE and its nearest serving SBS that has cached content $l$, $P_\mathrm{scd}$ is reformulated as
\begin{align}
P_\mathrm{scd}&=\sum_{l=1}^{L}p_l \mathbb{E}_{\left|X_l\right|}\left\{{\rm Prob}\left(\gamma_l\ge\gamma\bigg\vert \left|X_l\right|=x\right)\right\}\nonumber\\
&=\sum_{l=1}^{L}p_l\int_0^\infty {\rm Prob}\left(\gamma_l\ge\gamma\right)f_{\left|X_l\right|}\left(x\right)\mathrm{d}x,
\end{align}
where $f_{\left|X_l\right|}\left(x\right)$ is the PDF of $\left|X_l\right|$, given by~~\cite{F4_zhu2018content}
\begin{align}
f_{\left|X_l\right|}\left(x\right)=2\pi q_l\mu_\mathrm{S}x\exp\left(-\pi q_l\mu_\mathrm{S}x^2\right).\label{eq-pdf-x}
\end{align}
Therefore, the SCDP can be derived as
\begin{align}
P_\mathrm{scd}\hspace{-1mm}=\hspace{-1mm}\sum_{l=1}^{L} p_l \underset{P_{\mathrm{s},l}}{\underbrace{2\pi \mu_\mathrm{S} q_l\hspace{-1mm}\int_0^\infty\hspace{-1mm}\left[1-F_{g^{(l)}_\mathrm{FAS}}\left(\frac{\gamma\sigma^2_\eta}{P\beta}x^{\alpha}\right)\right]\hspace{-1mm}x\mathrm{e}^{-\pi q_l\mu_\mathrm{S}x^2}\mathrm{d}x}}.\label{eq-scdp}
\end{align}
Solving the integral~\eqref{eq-scdp} is mathematically intractable but it can be solved numerically. By setting the simulation parameters as $L=100$, $K=10$, $\textcolor{black}{A=4},$  $\mu_\mathrm{S}=10^{-2}~{\rm m}^{-2}$, $P=-30~{\rm dBm}$, $\sigma^2=-60~{\rm dBm}$, $\alpha=3$, $\beta=1$, $\gamma=0~{\rm dB}$, $\zeta=1$, and $T_0=1~{\rm ms}$, the SCDP and CDD are analyzed by the results in Figs.~\ref{fig-scd-eta} and~\ref{fig-scd-q} for various scenarios.


Fig.~\ref{fig-scd-eta} illustrates the impact of the SNR threshold $\gamma$ on the SCDP for $W$ and $N$, considering the scenario in which the caching placement probability is $q_l=1$. It can be seen that compared to the FPA system, deploying an FAS at the UE significantly enhances transmission reliability. \textcolor{red}{This is attributed to the spatial adaptability of FAS, which enables dynamic selection of antenna positions that optimize the instantaneous channel gain and reduce blockage effects, especially in dense deployments.} Moreover, simultaneously increasing both $W$ and $N$ improves the SCDP, as it balances the spatial correlation between antenna ports while enhancing channel capacity, diversity gain, and spatial multiplexing. \textcolor{red}{However, beyond a certain point, increasing $N$ may lead to a saturation of performance gains due to stronger spatial correlation between antenna ports and increased hardware complexity, which suggests a practical trade-off in FAS design.}
Fig.~\ref{fig-scd-q} illustrates the impact of FAS deployment on caching placement. The results show that SCDP increases monotonically with $q_l$ for both FAS and FPA, as higher caching probability enhances content accessibility. \textcolor{red}{Yet, the performance gain with FAS is more substantial, as its spatial adaptability helps establish stronger UE–SBS links, particularly when the cached content is available nearby, thereby maximizing the benefit of high $q_l$.} \textcolor{red}{ Moreover, increasing $\mu_\mathrm{S}$, which reflects the density of SBSs, improves the SCDP by reducing the average distance between UEs and their serving SBSs, thereby enhancing the overall SNR. In high $\mu_\mathrm{S}$ regimes, FAS systems benefit even more due to their ability to dynamically align with the strongest nearby SBS, leading to lower path loss and higher link quality.} \textcolor{red}{These observations highlight that deploying FAS not only improves physical-layer reliability but also amplifies the effectiveness of content placement strategies, indicating the potential for joint optimization of caching and antenna adaptation policies in future network designs.}

\begin{figure}[!t]
\centering
\includegraphics[width=\linewidth]{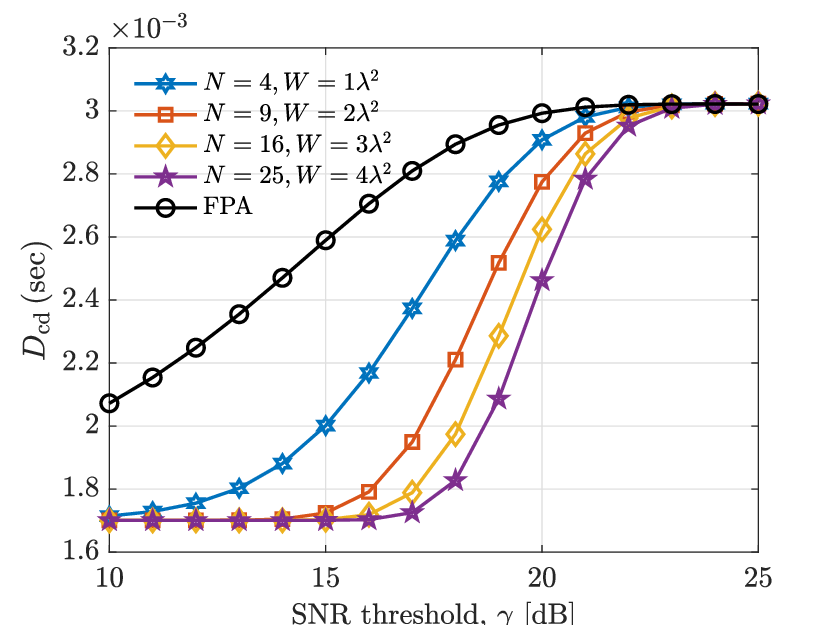}
\caption{CDD versus SNR threshold $\gamma$ when $q_l=1$ and $M=3$.}\vspace{-2mm}\label{fig-cd-eta}
\end{figure}
\begin{figure}[!t]
\centering
\includegraphics[width=\linewidth]{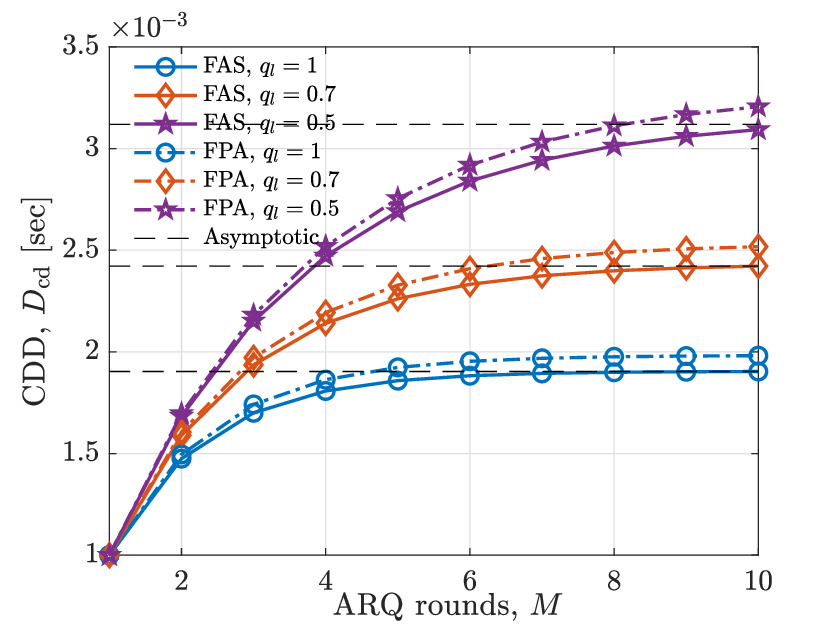}
\caption{CDD versus ARQ rounds $M$ when $N=3\times 3$, $W=1\lambda \times 1\lambda$.}\vspace{-2mm}\label{fig-cd-m}
\end{figure}

The ARQ-based protocol is considered, where an FAS-equipped UE requests content from the nearest SBS, which retransmits up to $\color{blue}M_{\rm re}$ times until successful delivery. Upon success, the SBS receives a one-bit acknowledgment; otherwise, a negative acknowledgment is sent. Each ARQ round lasts $T_0$, and an outage occurs if delivery fails after $\color{blue}M_{\rm re}$ attempts. Thus, the CDD is mathematically obtained as
\begin{align}\notag
D_\mathrm{cd}&=T_0+T_0\left(1-P_{\mathrm{s},l}\right)+\dots+T_0\left(1-P_{\mathrm{s},l}\right)^{{\color{blue}M_{\rm re}}-1}\\
&=T_0\frac{1-\left(1-P_{\mathrm{s},l}\right)^{\color{blue}M_{\rm re}}}{P_{\mathrm{s},l}},\label{ref-proof2}
\end{align}
where $P_{\mathrm{s},l}$ has been defined in~\eqref{eq-scdp}. Moreover, by considering the high ARQ rounds regime, i.e., ${\color{blue}M_{\rm re}}\rightarrow\infty$, the CDD in~\eqref{ref-proof2} can be derived as $D_\mathrm{cd}^\infty={T_0}/{P_{\mathrm{s},l}}$.

Fig.~\ref{fig-cd-eta} highlights the impact of FAS on CDD under varying SNR thresholds. FAS-equipped UEs experience significantly lower delays compared to FPA-based UEs. For instance, at $\gamma=15~{\rm dB}$, a UE with FAS ($N=3\times 3$, $W=\sqrt{2}\lambda\times\sqrt{2} \lambda$) achieves a CDD of approximately $1.72~{\rm ms}$, while a FPA-based UE faces a delay of about $2.6~{\rm ms}$, representing a reduction of $33.85\%$. Fig.~\ref{fig-cd-m} examines the effect of $\color{blue}M_{\rm re}$ on CDD. As $\color{blue}M_{\rm re}$ increases, delays rise due to more retransmission attempts. Additionally, lower caching probability results in higher delays, as the likelihood of the requested content being available at the nearest SBS decreases. Notably, FAS deployment consistently outperforms FPA system in these scenarios, delivering content with lower delays, even as $\color{blue}M_{\rm re}$ and $q_l$ fluctuate.

\section{Challenges and Open Issues}\label{sec:challenges}
The investigation into FAS has stimulated strong research interest. However, substantial efforts are required to facilitate the practical application of FAS. Based on our prior discussions, the following challenges and open issues lie ahead.

\subsection{System Modeling and Electromagnetic Compliance}
FAS necessitates precise electromagnetic models to accommodate dynamic changes in shape, position, material and other properties. Existing models struggle to achieve an appropriate balance between computational complexity and real-world accuracy. Moreover, the increasing operating frequencies necessitate the adoption of a near-field model. Therefore, it is imperative to explore accurate models for FAS's ultra-high spatial resolution in both far-field and near-field contexts.

\subsection{Channel Estimation and Management}
The dynamic reconfiguration of FAS mandates continuous channel estimation, increasing overhead in densely populated networks. Advanced technologies such as AI and machine learning approaches can potentially minimize the required channel acquisitions. Nonetheless, the number of observation ports required to accurately recover the CSI for large-scale configurations remains high. The ongoing reduction of CSI overhead without compromising performance continues to be a significant concern to reveal the capability of FAS.

\subsection{Hardware limitations}
The state-of-the-art channel models for FAS mainly emphasize theoretical frameworks, with empirical channel models remaining notably scarce. The development of an empirical channel model necessitates numerous channel measurements, a process that hinges on the maturity of FAS hardwares. Current FAS devices remains in its infancy, with existing prototypes inadequately developed for practical applications, thereby facing durability and miniaturization issues.

\subsection{Green FAS}
Given the rising carbon emissions resulting from excessive power consumption in wireless communication networks. EE has become a critical performance metric for future wireless infrastructures. However, the evaluation on the EE in FAS focuses on single-cell environment. Furthermore, current models consider circuit power as constant, while dynamic reconfiguration of FAS may lead to increased power demands. Consequently, a thorough analysis of the realistic EE performance of FAS is warranted, taking into account detailed circuit power dynamics in multi-cell scenarios.

\subsection{Security and Privacy}
The reconfigurability of FAS may introduce new vulnerability, including potential scenarios for beam-hijacking attacks. Robust encryption methodologies and dynamic beam-nulling techniques are still in the development stages. Furthermore, the spatial adaptability of FAS may inadvertently expose user location data, requiring anonymization frameworks. The physical layer vulnerability and the privacy issue beyond FAS need careful treatment and further investigations.

\subsection{Standardization and Deployment}
The high costs associated with the development of FAS infrastructure present a significant challenge. Should FAS be employed at the transmitter side to enhance DoF, necessary modifications to the feedback channel will be required. {\color{blue} New CSI feedback mechanisms and integration challenges with existing hybrid automatic repeat request (HARQ) protocols should be examined. Furthermore, the early stage of FAS development limits practical channel measurements, leading to a scarcity of benchmark datasets for standardized FAS channel models.} Nevertheless, as the integration of more antenna ports and efficient MIMO remains a focal topic of standardization, FAS  presents a promising solution to satisfy these requirements contingent upon appropriate design considerations.

\subsection{New Application Scenarios}
The introduction of FAS is driving rapid transformations within wireless networks. Due to its extensive DoF, it can be integrated into various emerging technologies to further enhance performance. For instance, FAS-enabled edge networks may improve transmission rates from the edge devices to servers, thereby improving computational capabilities. Numerous opportunities exist in new scenarios that extend beyond the scope of this discussion. Thorough investigation in the future will be crucial for the maturation of this technology.

\section{Conclusion}\label{sec:conclusion}
This paper provided a contemporary survey on the pertinent subject of FAS, encompassing key application scenarios, fundamental physical-layer techniques, and relevant networking technologies. The application scenarios were categorized into three principal types: single-user HomoNet, multi-user HomoNet, and HetNets. Subsequently, the fundamental aspects of FAS were presented, with particular focus on channel modeling and channel estimation as crucial components for understanding FAS behavior. The paper further summarized transmission models, theoretical analyses, link-level simulation results, and their variants for both single-user FAS and multi-user FAMA. At the time of this writing, there is growing interest in associated networking solutions. To this end, the discussion included techniques for QoS enhancement in FAS networks and strategies for power allocation under QoS constraints. In addition, this paper addressed content placement strategies in FAS-based HetNets, which can improve system efficiency by reducing delivery delays. Further research into networking techniques for FAS can be pursued from multiple perspectives. In summary, this marks an exciting period for researchers engaged in the study of FAS.

%%%%%%%%%%%%%%%%%%%%%%%%%%%%%%%%%%%%%%%%%%%%%%%%%%%%%%%%%%%%%%%%%%%%%%%%%%%%%%%%%%%%%%%%%%%%%%%%%%%%%%%%%%%%%%%%%%%%%%%%%%%%%%%%%%%%%%%%%%%%%%%%%%%%%%%%%%%%%%%%
\bibliographystyle{IEEEtran}
%  \bibliography{Reference}

\end{document}